\documentstyle[12pt,preprint,aps,floats,epsf,graphicx,epsfig]{revtex}
\tighten

\begin{document}

\preprint{
\noindent
\hfill
\begin{minipage}[t]{6in}
\begin{flushright}
CERN-TH/2001-136\\
BUTP-01/12\\
ZU-TH 15/01\\
hep-ph/0105292  \\
\vspace*{1.0cm}
\end{flushright}
\end{minipage}
}

\draft

\title{\Large \bf Bounds on Flavour Violating Parameters\\ in 
Supersymmetry\footnote{Work partially supported by 
Schweizerischer Nationalfonds}}

\author{ {\bf Thomas ~Besmer\,} (a), \,{\bf Christoph ~Greub\,} (b), \,  
  {\bf Tobias ~Hurth\,} (c) }

\vspace{1cm}

\address{\it 
 \vspace{1cm}
a \,  
Institute for Theoretical Physics, University of  Zurich,
 CH--8057 Zurich, Switzerland\\
 \vspace{0.5cm}
b \, 
 Institute for Theoretical  Physics, University of Berne, 
 CH--3012 Berne, Switzerland\\ 
 \vspace{0.5cm}
c \, 
Theory Division, CERN, CH--1211 Geneva 23, Switzerland}

\maketitle

\vspace*{1truecm}
\begin{abstract}
We investigate the constraints 
on the flavour violating parameters 
from the decay $B \rightarrow X_s \gamma$, taking 
into account the interplay of the various sources of flavour violation in
the unconstrained MSSM. 
We present a systematic leading logarithmic QCD analysis of 
these model-independent constraints, including contributions from
gluinos, neutralinos, charginos, charged Higgs bosons  
and interferences between them.
We show that two simple combinations of elements
of the down squark mass matrix are stringently bounded 
over large parts of the parameter space 
where only weak assumptions on the hierarchical
structure of the squark mass matrices are made.
We also briefly analyse up to which values SUSY contributions, 
compatible with 
$B \rightarrow X_s \gamma$,
can enhance the Wilson coefficient $C_8(m_W)$, which 
plays an important role in the phenomenology of charmless hadronic 
$B$ decays.

\end{abstract}

\vfill

%\today

\newpage

\setlength{\parskip}{1.2ex}

\section{Introduction}
\label{intro}

\vspace{0.5 cm}
Today supersymmetric models are given priority in the  search for 
new physics beyond the standard model  (SM). 
This is primarily suggested by theoretical arguments
related to the hierarchy problem. 
Supersymmetry eliminates the sensitivity for the
highest scale in the theory and, thus, stabilizes the low energy theory.

Flavour changing neutral current (FCNC) processes
provide crucial guidelines for supersymmetry model building. 
In the so-called unconstrained minimal supersymmetric standard model
(uMSSM) there are new sources for FCNC transitions. 
Besides
the Cabibbo--Kobayashi--Maskawa (CKM)-induced contributions,
there are  generic supersymmetric  contributions induced by 
flavour mixing  in the squark mass matrices.
The structure of the uMSSM does not explain the suppression of 
FCNC processes, which is observed in experiments;
this is the crucial point of the well-known supersymmetric flavour problem.
Within the framework of the MSSM there are at present three favoured 
supersymmetric models that solve the supersymmetric 
flavour problem by a specific mechanism through which the sector of 
supersymmetry breaking 
communicates with the sector accessible to experiments: in the 
minimal supergravity model (mSUGRA)~\cite{MSUGRA},  
supergravity is the corresponding
mediator; in the other two models, this is achieved by gauge 
interactions~\cite{GMSB}
 and by anomalies~\cite{AMSB}. Furthermore, there are other 
classes of models in which the flavour problem is solved by particular 
flavour symmetries~\cite{FLAVOUR}.

Flavour violation thus originates
from the interplay  between the dynamics of flavour and the mechanism of  
supersymmetry breaking.  FCNC processes
therefore yield important (indirect) information on the construction 
of supersymmetric extensions of the SM and can contribute 
to the question of which mechanism ultimately breaks supersymmetry. 
The experimental measurements of the rates for these processes, or the
upper limits set on them, impose in general a reduction of the large
number and size of parameters in the soft supersymmetry-breaking terms
present in these models. Among these processes, those involving
transitions between first- and second-generation quarks, namely FCNC
processes in the $K$ system, are considered as the most formidable
tools to shape viable supersymmetric flavour models. Moreover, the
tight experimental bounds on some flavour-diagonal transitions, such
as the electric dipole moment of the electron and of the neutron, as
well as $g-2$, help constraining soft terms inducing chirality
violations.

Among neutral flavour transitions involving the third generation, 
the rare decay $B \rightarrow X_s \gamma$
is at present the most important one, as it is
the only inclusive mode which is already 
measured~\cite{BSGMEASURE}.
The theoretical SM prediction,  up to  next-to-leading 
logarithmic (NLL) precision~\cite{NLL}
for its branching ratio,
is in agreement with the experimental data.
Although the experimental errors are still rather large, this 
decay mode already 
allows for theoretically clean and rather stringent constraints on 
the parameter space of various extensions of the 
SM (see for example~\cite{NLLBEYOND}).

Once more precise data from the B factories are available, 
this decay will undoubtedly gain efficiency
in selecting the viable regions of the parameter space in the above
classes of models; it may also help discriminating among the models that 
will be proposed by then.

In this paper we present a model-independent 
analysis of the decay
 $B \rightarrow X_s\,\gamma$, based on a LL-QCD calculation,
where contributions from $W$-bosons, charged Higgs bosons, 
charginos, neutralinos
and gluinos are systematically included. 
Former analyses in the unconstrained MSSM neglected QCD corrections
and only used 
the gluino contribution to saturate the experimental bounds. 
Technically, the so-called mass insertion approximation (MIA) 
was  used where the off-diagonal elements of the squark mass matrices
are taken to be small and their higher powers neglected. 
As a consequence of this single insertion approximation, 
no correlations between different sources of flavour 
violation were taken into account. 
In this way, one arrived at  'order-of-magnitude bounds' on the 
soft parameters~\cite{GGMS,DNW,HKT}.
In~\cite{BGHW}, the sensitivity of the bounds on the down squark
mass matrix to radiative QCD corrections was analysed
including the SM and the gluino contributions.
The aim of the present paper is  to extend this analysis
to include the contributions from charged Higgs bosons, charginos
and neutralinos and their interference effects, and even more important,
the effects that result when several flavour violating parameters,
i.e. several off-diagonal elements in the squark mass matrices, 
are switched on simultaneously. We anticipate that two simple combinations
of matrix elements of the down squark mass matrix remain rather stringently
bounded over large parts of the parameter space, in a general scenario where only relatively weak assumptions 
on pattern of the squark mass matrices are made. 

Since there are different contributions to this
decay, with different numerical impact on its rate, some of these
flavour-violating terms may turn out to be poorly constrained. Thus, 
given the generality of such a calculation, it is 
convenient to rely on the mass eigenstate 
formalism, which remains valid even when some of the intergenerational mixing 
elements are large, and not to use the approximate mass insertion 
method, where the off-diagonal squark mass matrix elements are taken 
to be small and their higher powers neglected. In the latter approach
the reliability of the approximation can only be checked a posteriori.

Finally, we note that the off-diagonal elements of the squark mass matrices 
can get constraints on completely different grounds, namely from the 
requirement of the absence of charge and colour breaking minima 
as well as  from directions in the scalar potential which are unbounded
from below (see \cite{Casas} for a more detailed discussion).
However, these bounds correspond to sufficient, but not necessary conditions 
for the stability of the standard vacuum,  because it 
is possible  that we live in a  metastable vacuum, whose lifetime is 
longer than the  age of the universe \cite{Kusenko}.

The paper is organized as follows: In section 2, we discuss
the framework for the calculation of the branching ratio
for $B \rightarrow X_s \gamma$.
In section 3, we briefly recall the sources of flavour violation, encoded
in the squark mass matrices.  In section 4, we present 
the phenomenological analysis on the bounds on the flavour violating 
parameters. In section 5,  
we briefly explore  up to which values SUSY contributions,  allowed 
by \mbox{$B \rightarrow X_s \gamma$},
can enhance the Wilson coefficient $C_8(m_W)$, which 
plays an important role in the phenomenology of charmless hadronic 
$B$ decays. In section 6 we give a short summary.
In appendix A1, we state our conventions, while in appendix A2 we list
the Wilson coefficients at the matching scale. 
\section{Framework for calculating $B \rightarrow X_s
 \gamma$}
\label{framework}
\subsection{Hamiltonians}
In the SM, rare $B$-meson decays are induced by one-loop diagrams in which $W$
bosons and up-type quarks propagate.  
The most important corrections to the decay amplitude for $b \to s \gamma$ 
are due to exchanges of gluons,
which give rise to powers of the factor
$L=\log(m_b^2/m_W^2)$. It turns out that each of these logarithms 
is accompanied by at least one factor of $\alpha_s$.  
Since the two scales $m_b$ and
$m_W$ are far apart, $L$ is a large number and these terms need to be
resummed: at the leading logarithmic (LL) order, powers of $\alpha_s L$ 
are resummed; at the next-to-leading (NLL) order, also the 
terms of the form $\alpha_s \, (\alpha_s L)^N$ are systematically  resummed.  
Thus, the
contributions to the decay amplitude are classified according to
\begin{equation}
\quad (LL): \quad \quad G_F \, (\alpha_s L)^N, \quad 
 \quad (NLL):  \quad       G_F \, \alpha_s (\alpha_s L)^N, \quad (N=0,1,...)  
\label{terms}
\end{equation}
where $G_F$ is the Fermi constant. The resummation of 
these corrections is usually achieved by making use of
the formalism of effective Hamiltonians, combined with renormalization
group techniques. The effective Hamiltonian 
${\cal H}_{eff}^{W}$, 
obtained by integrating out the top-quark and
the $W$ boson, can be written as 
\begin{equation}
 {\cal H}_{eff}^{W} = 
- \frac{4 G_F}{\sqrt{2}} K_{tb}^{\phantom{\ast}} K_{ts}^\ast
  \sum_i C_i(\mu) {\cal O}_i(\mu) \,. 
\label{weffham}
\end{equation}
The Wilson
coefficients $C_i$ contain all dependence on the heavy degrees of
freedom, whereas the operators ${\cal O}_{i}$ depend on light
fields only. 
The operators relevant to $b \to s \gamma$
read
\begin{equation}
\begin{array}{llll}
{\cal O}_{1}                 \,= &\!
(\bar{s} \gamma_\mu T^a P_L c)\,  (\bar{c} \gamma^\mu T_a P_L b)\,, 
               &         \\[1.2ex]
{\cal O}_{2}                 \,= &\!
(\bar{s} \gamma_\mu P_L c)\,  (\bar{c} \gamma^\mu P_L b)\,,    
               &         \\[1.2ex]
{\cal O}_{3}                 \,= &\!
(\bar{s} \gamma_\mu P_L b) \sum_q (\bar{q} \gamma^\mu q)\,,    
               &        \\[1.2ex]
{\cal O}_{4}                 \,= &\!
(\bar{s} \gamma_\mu T^a P_L b) \sum_q (\bar{q} \gamma^\mu T_a q)\,,
               &        \\[1.2ex]
{\cal O}_{5}                 \,= &\!
(\bar{s} \gamma_\mu \gamma_\nu \gamma_\rho P_L b) 
 \sum_q (\bar{q} \gamma^\mu \gamma^\nu \gamma^\rho q)\,,        
               &        \\[1.2ex]
{\cal O}_{6}                 \,= &\!
(\bar{s} \gamma_\mu \gamma_\nu \gamma_\rho T^a P_L b) 
 \sum_q (\bar{q} \gamma^\mu \gamma^\nu \gamma^\rho T_a q)\,,      
               &  \\[1.2ex]
{\cal O}_{7}                 \,= &\!
  \displaystyle{\frac{e}{16\pi^2}} \,{\overline m}_b(\mu) \,
 (\bar{s} \sigma^{\mu\nu} P_R b) \, F_{\mu\nu}\,,      \\[2.0ex]    
{\cal O}_{8}                 \,= &\!
  \displaystyle{\frac{g_s}{16\pi^2}} \,{\overline m}_b(\mu) \,
 (\bar{s} \sigma^{\mu\nu} T^a P_R b)
     \, G^a_{\mu\nu}\,.                               
\label{smbasis}                                        
\end{array}
\end{equation}
The matrices $T^a$ ($a=1,...,8$) are $SU(3)$ colour generators and 
$P_{L,R}$ are left- and right-handed projection operators;
$e$ and $g_s$ denote the electromagnetic and the strong coupling
constants, respectively. Note that the $b$-quark mass is the relevant parameter
that governs the chirality flip in the SM dipole operators
${\cal O}_{7} $ and ${\cal O}_{8} $. All eight operators are of dimension six. 
We anticipate that this is in
contrast with the dipole operators induced by gluinos,
where the helicity flip can be generated by the gluino mass instead of the
$b$-quark mass, as we will see in more detail later. As a consequence, these
dipole operators are effectively of dimension five.

A consistent SM calculation for 
$B \to X_s \gamma$ 
at LL (or NLL) precision requires three steps:
%\\{\it 1)}
\begin{itemize}
\item[{\it 1)}] 
a matching calculation of the full SM theory 
with the effective theory at the scale $\mu=\mu_W$ 
to order $\alpha_s^0$ (or $\alpha_s^1$) for the Wilson coefficients, 
where  $\mu_W$ denotes a scale of order $m_W$ or $m_t$;
%\\{\it 2)}
\item[{\it 2)}]  
a renormalization group treatment of the Wilson coefficients
using the anomalous-dimension matrix to order $\alpha_s^1$ 
(or $\alpha_s^2$);
%\\{\it 3)}
\item[{\it 3)}]   
a calculation of the operator matrix elements at the scale 
$\mu = \mu_b$  to order $\alpha_s^0$ (or $\alpha_s^1$), 
where $\mu_b$ denotes a scale of order $m_b$.
\end{itemize}

In supersymmetric models there are further contributions  
to the FCNC processes studied 
in this paper, i.e. 
the exchange of charged Higgs bosons 
and up-type quarks; of charginos and up-type squarks; 
of neutralinos and down-type squarks; and of gluinos and down-type
squark. They all lead to
$|\Delta(B)|=|\Delta(S)|=1$ effective magnetic and chromomagnetic
operators (of ${\cal O}_7$-type, ${\cal O}_8$-type)
and also to new four-quark operators.

Taking into account operators up to dimension six only, the
effects of
charginos, neutralinos and charged Higgs bosons can be matched
onto the usual SM magnetic and chromomagnetic operators 
${\cal O}_{7}$ and ${\cal O}_{8}$ and their primed counterparts
\begin{equation}
\begin{array}{llll}
 {\cal O}_{7}^\prime                 \,= &\!
  \displaystyle{\frac{e}{16\pi^2}}\,{\overline m}_b(\mu)
 \,(\bar{s} \sigma^{\mu\nu} P_L b) \, F_{\mu\nu}\,,   
                                        &  \quad 
 {\cal O}_{8}^\prime              \,= &\!
  \displaystyle{\frac{g_s}{16\pi^2}} \,{\overline m}_b(\mu)
 \,(\bar{s} \sigma^{\mu\nu} T^a P_L b) \, G^a_{\mu\nu}\,.
\label{smmagnopw2}
\end{array}
\end{equation}
We would like to stress that we do not work in the mSUGRA scenario.
Therefore, a lot of the relations deduced in~\cite{BBMR} 
do not hold anymore. The most important 
thing to notice is that in a general SUSY 
framework, there is no connection between the Yukawa couplings in the
superpotential and the corresponding trilinear term in the soft potential. 
However, the contributions from charginos
and charged Higgs bosons to the Wilson coefficients of the primed
operators vanish also in the more general unconstrained MSSM
if $m_s$ is put to zero
(see eqs. (\ref{C7'}) and (\ref{C8'})). 
This implies that for physical values of  $m_s$ the chargino- and the charged
Higgs boson contributions to the primed operators
are small in scenarios in which
$\tan \beta$ does not take extreme values. 
The neutralino contributions to both, the primed and unprimed
operators are also unimportant, because their Wilson coefficients
involve those entries of the squark mixing  matrices $\Gamma_{DL}$ 
and $\Gamma_{DR}$, which also induce gluino contributions;
the latter, which are proportional to $g_s^2$ therefore dominate
the neutralino contributions which are proportional to $g_2^2$.
We also found numerically that the neutralino contibutions are indeed
rather inessential.

The fact that the operators generated by charged Higgs bosons,
charginos and neutralinos can be matched  onto the SM operators
and their primed counterparts
implies that the terms that get resummed at LL, 
show the same pattern as those listed in
eq.  (\ref{terms}); the Fermi constant appearing there  
is obviously replaced by a specific supersymmetric parameter for the
chargino and neutralino contributions.

Matters are somewhat different for the gluino contribution 
${\cal H}_{eff}^{\tilde{g}}$, as worked out in detail 
in \cite{BGHW}. At the amplitude level, 
terms of the form
\begin{itemize}
\item[] \quad $(LL)$:
 $\quad \quad \alpha_s \, (\alpha_s L)^N,$ \quad $(NLL)$: 
 $\quad \alpha_s \, \alpha_s (\alpha_s L)^N$,  
\quad $(N=0,1,...)$ 
\end{itemize}
are resummed, respectively at the leading and next-to-leading order.  
While ${\cal H}_{eff}^{\tilde{g}}$ is unambiguous, it is a matter of
convention whether the $\alpha_s$ factors
should be put into the
definition of operators or into the Wilson coefficients.
We follow the framework developed in ref. \cite{BGHW},
where the distribution of the $\alpha_s$ factors was 
done in such a way that the anomalous dimension matrix systematically starts
at order $\alpha_s$. 
We write the effective Hamiltonian  
${\cal H}_{eff}^{\tilde{g}}$ in the form
\begin{equation}
 {\cal H}_{eff}^{\tilde{g}} = 
 \sum_i C_{i,\tilde{g}}(\mu) {\cal O}_{i,\tilde{g}}(\mu)  +
 \sum_i \sum_q C_{i,\tilde{g}}^q(\mu) {\cal O}_{i,\tilde{g}}^q(\mu) \,. 
\label{geffham}
\end{equation}
The index $q$ runs over all light quarks
$q=u,d,c,s,b$. Among the operators contributing to the first part,
there are
dipole operators in which the  chirality flip
is induced by the $b$-quark mass:
\begin{equation}
\begin{array}{llll}
{\cal O}_{7b,\tilde{g}}                 \,= &\!
   e \,g_s^2(\mu) \,{\overline m}_b(\mu) \,
 (\bar{s} \sigma^{\mu\nu} P_R b) \, F_{\mu\nu}\,,   
                                        &  %\quad 
{\cal O}_{7b, \tilde{g}}^{\prime}        \,= &\!
   e \,g_s^2(\mu) \,{\overline m}_b(\mu) \,
 (\bar{s} \sigma^{\mu\nu} P_L b) \, F_{\mu\nu}\,,     \\[2.0ex]    
%                                       &             \\
{\cal O}_{8b, \tilde{g}}                 \,= &\!
 g_s^3(\mu) \,{\overline m}_b(\mu) \,
 (\bar{s} \sigma^{\mu\nu} T^a P_R b)
     \, G^a_{\mu\nu}\,,                               
                                        &  %\quad 
{\cal O}_{8b, \tilde{g}}^\prime          \,= &\!
 g_s^3(\mu) \,{\overline m}_b(\mu) \,
 (\bar{s} \sigma^{\mu\nu} T^a P_L b)
     \, G^a_{\mu\nu}\,.
\label{gmagnopb} 
\end{array}
\end{equation}
As discussed in \cite{BGHW}, there are also gluino-induced 
operators where the chirality violation is signalled by 
the charm quark mass (obtained by replacing 
${\overline m}_b(\mu)$ by ${\overline m}_c(\mu)$) and operators where the 
chirality flip is induced by the gluino mass. The latter read
\begin{equation}          
\begin{array}{llll}  
{\cal O}_{7\tilde{g},\tilde{g}}         \,= &\!
  e \,g_s^2(\mu) \,
 (\bar{s} \sigma^{\mu\nu} P_R b) \, F_{\mu\nu}\,,    
                                        &  \quad 
{\cal O}_{7\tilde{g},\tilde{g}}^\prime  \,= &\!
  e \,g_s^2(\mu) \,
 (\bar{s} \sigma^{\mu\nu} P_L b) \, F_{\mu\nu}\,,     \\
                                        &             \\[-1.3ex]           
{\cal O}_{8\tilde{g},\tilde{g}}         \,= &\!
 g_s^3(\mu) \,
 (\bar{s} \sigma^{\mu\nu} T^a P_R b)
     \, G^a_{\mu\nu}\,, 
                                        &  \quad 
{\cal O}_{8\tilde{g},\tilde{g}}^\prime  \,= &\!
 g_s^3(\mu) \,
 (\bar{s} \sigma^{\mu\nu} T^a P_L b)
     \, G^a_{\mu\nu}\,. 
\label{gmagnopg}                                     
\end{array} 
\end{equation}
At the LL-level, these operators could be absorbed into the 
operators given in eq. (\ref{gmagnopb}), when neglecting
the small mixings effects from the gluino-induced 
four-Fermi operators with scalar or tensor Lorentz structure.
However, as it is useful to
separate the contributions where the chirality flip is induced by 
$m_{\tilde{g}}$, we do not perform this absorption. 
Notice that the operators in eq. (\ref{gmagnopg}) have dimension 
{\it five}, while
the other operators are of 
dimension {\it six}.  We also stress that unlike the other 
supersymmetric 
contributions, the primed gluino-induced operators 
are {\it not} suppressed 
compared with the unprimed ones. This is in strong contrast with the
mSUGRA scenario, where the primed operators are stronlgy
suppressed~\cite{BBMR}.

The contributions to the second part in  eq. (\ref{geffham}) are
given by  four-quark operators with vector, scalar and tensor
Lorentz structure.
As shown in ref. \cite{BGHW}, the scalar and tensor operators
mix at  one loop into the six-dimensional magnetic and chromomagnetic ones.  
Therefore, they have to be included
in principle in a  LL calculation. As mentioned above, these mixings
are numerically small and therefore not very important in practice.   

For completeness we recall all Wilson coefficients  at the matching scale 
$\mu_W$ in appendix~\ref{Wilson}.  
The anomalous dimension matrix of the SM operators 
${\cal O}_1$--${\cal O}_8$ and the evolution of the corresponding
Wilson coefficients to the decay scale $\mu_b$ are well known and
can be found in~\cite{NLL}.
The evolution of the gluino-induced Wilson coefficients 
$C_{i,\tilde{g}}$ is given in ref.~\cite{BGHW}.

\subsection{Branching ratio}
The decay width for $B \rightarrow X_s \gamma$
to LL precision is given by
\begin{equation}
 \Gamma(B \to X_s \gamma) = 
 \frac{m_b^5 \, G_F^2 \, |K_{tb} K_{ts}^*|^2 
 \, \alpha}{32 \pi^4} \ 
 \left(\left\vert \hat{C}_7(\mu_b)  \right\vert^2 + 
       \left\vert \hat{C}'_7(\mu_b) \right\vert^2
 \right) \,,
\end{equation}
where the auxiliary quantities 
$\hat{C}_7(\mu_b)$ and $\hat{C}'_7(\mu_b)$ are defined as
\begin{eqnarray}
\hat{C}_7(\mu_b) & = &
C_7^{\rm{eff}}(\mu_b) -
\left[           C_{7b,\tilde{g}}(\mu_b) + 
 \frac{1}{m_b}   C_{7\tilde{g},\tilde{g}}(\mu_b) 
\right] \,  
 \frac{16 \sqrt{2} \pi^3 \alpha_s(\mu_b)}{G_F \, K_{tb} K_{ts}^*} \, , 
 \nonumber  \\ 
\hat{C}'_7(\mu_b) & = &
C_7'(\mu_b) \, 
 -  \left[           C'_{7b,\tilde{g}}(\mu_b) + 
 \frac{1}{m_b}   C'_{7\tilde{g},\tilde{g}}(\mu_b) 
\right] \,
 \frac{16 \sqrt{2} \pi^3 \alpha_s(\mu_b)}{G_F \, K_{tb} K_{ts}^*} \, ,
\label{c7hat}
\end{eqnarray}
where  
\begin{equation}
C_7^{\rm{eff}}(\mu_b) =  C_7(\mu_b) 
                        - \frac{1}{3}  C_3(\mu_b)
                        - \frac{4}{9}  C_4(\mu_b)
                        - \frac{20}{3} C_5(\mu_b)
                        - \frac{80}{9} C_6(\mu_b) \, .
\end{equation}
Note that we have neglected the small contributions from the operators
${\cal O}_{7c,\tilde{g}}$ and 
${\cal O}_{7c,\tilde{g}}^\prime$.  
The branching ratio
can be expressed as 
\begin{equation}  
 {\rm BR}(B \to X_s\gamma) =
 \frac{\Gamma(B \to X_s \gamma)}{\Gamma_{SL}} \, {\rm BR}_{SL} \, ,
\label{bratio}
\end{equation}  
where ${\rm BR}_{SL}=(10.49 \pm 0.46)\%$ is 
the measured semileptonic branching
ratio. To the relevant order in $\alpha_s$, the semileptonic decay width
is given by:
\begin{equation}
\Gamma_{SL} =
 \frac{m_b^5 \, G_F^2 \, |K_{cb}|^2}{192 \pi^3}
 \, g\left(\frac{m_c^2}{m_b^2}\right) \,,
\end{equation}
where the phase-space function $g(z)$ is 
$g(z) = 1 - 8z + 8 z^3 - z^4 - 12 z^2 \log z$.

Note that there is no systematic distinction
between the pole mass $m_b$ and the 
corresponding running mass normalized at the scale $\mu_b$
in the LL approximation. 
To be specific, the mass parameters are always treated as 
pole masses in our numerical analysis. 

\section{Squark Mass Matrices as New Sources of Flavour Violation} 
\label{SMM}
As advocated in the introduction, the aim of this paper is to 
provide a phenomenological analysis of the constraints on the
flavour violating parameters in supersymmetric models with the most 
general soft terms in the squark mass matrices. 
As explained there, we work in the mass eigenstate formalism, 
which remains valid (in contrast to the mass insertion approximation) 
when the intergenerational mixing elements are not small.

A specification of the squark mass matrices usually starts in the
super-CKM basis, in which the superfields are rotated in such a way that
the mass matrices of the quark fields are diagonal.
In this basis, the $6\times 6$ squared-mass matrix for the $d$-type squarks 
has the form 
\begin{equation}
{\cal M}_d^2 \equiv  \left( \begin{array}{cc}
  m^2_{\,d,\,LL} +F_{d\,LL} +D_{d\,LL}           & 
                 \left(m_{\,d,\,LR}^2\right) + F_{d\,LR} 
                                                     \\[1.01ex]
 \left(m_{\,d,\,LR}^{2}\right)^{\dagger} + F_{d\,RL} &
             \ \ m^2_{\,d,\,RR} + F_{d\,RR} +D_{d\,RR}                
 \end{array} \right) \,.
\label{massmatrixd}
\end{equation}
For the $u$-type squarks we have
\begin{equation}
{\cal M}_u^2 \equiv  \left( \begin{array}{cc}
  m^2_{\,u,\,LL} +F_{u\,LL} +D_{u\,LL}           & 
                 \left(m_{\,u,\,LR}^2\right) + F_{u\,LR} 
                                                     \\[1.01ex]
 \left(m_{\,u,\,LR}^{2}\right)^{\dagger} + F_{u\,RL} &
             \ \ m^2_{\,u,\,RR} + F_{u\,RR} +D_{u\,RR}                
 \end{array} \right) \,.
\label{massmatrixu}
\end{equation}
In this basis, the $F$ terms  (stemming from the superpotential) 
in the $6 \times 6$
mass matrices ${\cal M}^2_f$ ($f=u,d$) are diagonal 
$3 \times 3$ submatrices, reading
$(F_{f\,RL}=F_{f\,LR}^\dagger)$
\begin{equation}
(F_{d\,LR})_{ij} =  -\mu (m_{d,i} \tan \beta)\, 
 {{\mathchoice {\rm 1\mskip-4mu l} {\rm 1\mskip-4mu l}
{\rm 1\mskip-4.5mu l} {\rm 1\mskip-5mu l}}}_{ij}
\,, \,\,\, 
(F_{u\,LR})_{ij} =  -\mu (m_{u,i} \cot \beta)\,  {{\mathchoice {\rm 1\mskip-4mu l} {\rm 1\mskip-4mu l}
{\rm 1\mskip-4.5mu l} {\rm 1\mskip-5mu l}}}_{ij} \, ,
\label{FFterm}
\end{equation}
\begin{equation}
(F_{d\,LL})_{ij} =  m^2_{d\,i}\,  {{\mathchoice {\rm 1\mskip-4mu l} {\rm 1\mskip-4mu l}
{\rm 1\mskip-4.5mu l} {\rm 1\mskip-5mu l}}}_{ij}  \, , \,\,\, 
(F_{u\,LL})_{ij} =  m^2_{u\,i}\,  {{\mathchoice {\rm 1\mskip-4mu l} {\rm 1\mskip-4mu l}
{\rm 1\mskip-4.5mu l} {\rm 1\mskip-5mu l}}}_{ij} \, .
\label{Fterm}
\end{equation}
Also the $D$-term contributions $D_{f\,LL}$ and $D_{f\,RR}$ 
to the squared-mass matrix are diagonal in flavour space: 
\begin{equation}
 D_{f\,LL,RR} =  \cos 2\beta \, m_Z^2 
   \left(T_f^3 - Q_f \sin^2\theta_W \right) 
{{\mathchoice {\rm 1\mskip-4mu l} {\rm 1\mskip-4mu l}
{\rm 1\mskip-4.5mu l} {\rm 1\mskip-5mu l}}}_3\,.
\label{dterm}
\end{equation}
Since present collider limits give indications that 
the squark masses are larger than those of the corresponding 
quarks, the largest entries in the squark mass matrices squared
must come from the soft potential, directly 
linked to the mechanism of supersymmetry breaking. 
These contributions, denoted in (\ref{massmatrixd}) and (\ref{massmatrixu}) 
by $m^2_{\,f,\,LL}$, $m^2_{\,f,\,RR}$ and $m^2_{\,f,\,LR}$, are 
in general not diagonal in the super-CKM basis.

Further comments are in order.
Because of $SU(2)_L$ invariance, 
$m^2_{\,u,\,LL}$ and $m^2_{\,d,\,LL}$ are related. In the super-CKM basis
this relation reads
$m^2_{\,u,\,LL} = K m^2_{d,LL} K^{\dagger}$,
where $K$ denotes the CKM matrix.
The off-diagonal $3 \times 3$ block matrix $ m_{\,f,\,LR}^2$ 
equals $A^{\,\ast}_{d} v_d $ for down-type and
$A^{\,\ast}_{u} v_u $ for up-type squarks
(the two vacuum expectation values are chosen to be real).
They arise from the trilinear terms in the soft potential,
namely
$A_{d,ij} H_d \,{\widetilde{D}}_i {\widetilde{D}}_j^{c}$ and   
$A_{u,ij} H_u \,{\widetilde{U}}_i {\widetilde{U}}_j^{c}$.
We stress that we do {\it not} assume the proportionality 
of these trilinear terms to the Yukawa couplings, as is
done in the mSUGRA model. 
Furthermore, differently from $ m^2_{\,f,\,LL}$
and $ m^2_{\,f,\,RR}$, the off-diagonal $3 \times 3$ matrix 
$m_{\,f,\,LR}^2$ is not hermitian. 

Because 
all neutral gaugino couplings are flavour diagonal
in the super-CKM basis
and the mixing in the charged gaugino coupling to quarks and squarks is
governed by the conventional CKM matrix, 
the flavour change through squark mass mixing is parametrized 
by the off-diagonal
elements of the soft terms 
$m^2_{f,LL}$, $m^2_{f,RR}$, $m^2_{f,LR}$ in this basis.

The diagonalization of the two $6 \times 6$ 
squark mass matrices ${\cal M}^2_d$ and ${\cal M}^2_u$  
yields the eigenvalues $m_{\tilde{d}_k}^2$ and $m_{\tilde{u}_k}^2$ 
($k=1,...,6$). The corresponding mass eigenstates, 
$\tilde{u}_{k}$ and $\tilde{d}_{k}$ ($k=1,...,6$) are 
related to the fields in the super-CKM basis, 
$\tilde{u}_{Lj}$, $\tilde{u}_{Rj}$
and $\tilde{d}_{Lj}$, $\tilde{d}_{Rj}$, 
($j=1,...,3$) as 
\begin{equation}
\tilde{u}_{L,R} = \Gamma^\dagger_{UL,R} \, \tilde{u} \,,
\hspace*{1truecm}
\tilde{d}_{L,R} = \Gamma^\dagger_{DL,R} \, \tilde{d} \,,
\label{qdiag}
\end{equation}
where the four matrices
$\Gamma_{UL,R}$ and $\Gamma_{DL,R}$ are $6 \times 3$ mixing
matrices. 
\section{Phenomenological Analysis}
\label{Pheno}
Our phenomenological analysis is based on 
a complete LL QCD calculation within the unconstrained MSSM;
it is done in two parts:

\begin{itemize}
\item 
In the first part, we try to derive bounds on the off-diagonal elements of the
squark mass matrices by 
switching on only {\it one} of these elements at a time. 
We include, however, 
all new physics contributions (chargino,
neutralino, charged Higgs bosons, gluino) in the analysis. 
We show that only those parameters get stringently
bounded by $B \to X_s \gamma$, which can generate
contributions to the five-dimensional gluino-induced dipole operators
${\cal O}_{7\tilde{g},\tilde{g}}$ and 
${\cal O}'_{7\tilde{g},\tilde{g}}$.

\item 
In the second part of our analysis we investigate whether 
the bounds obtained in the first part remain stable,
if {\it all} off-diagonal elements, which induce the 
decay $B \rightarrow X_s \gamma$, are varied simultaneously.
We anticipate that the bounds on the individual off-diagonal
elements get lost, because in this case various combinations of off-diagonal
elements can contribute (with opposite sign)
to the Wilson coefficients of the five-dimensional 
dipole operators. In the scenarios we discuss 
below, it is, however, possible to constrain certain simple combinations  
of off-diagonal elements of the down squark mass matrices, provided
$\tan \beta$ and $\mu$ are not very large.
\end{itemize}

\subsection{General comments}
In order to analyse the implications of $B \to X_s \gamma$
on the flavour violating soft  parameters in the squark mass
matrices, we choose some specific scenarios that are characterized
by the values of the parameters\\

\begin{equation}
\mu, \quad  M_{H^-}, \quad  \tan \beta, \quad  M_{\rm{susy}}, \quad 
 m_{\tilde{g}}. 
\label{susypar}
\end{equation}

We regard this as reasonable, because we expect that
these input parameters, which are unrelated to flavour physics,
will be fixed from flavour conserving observables in the next generations
of high energy experiments (provided low energy SUSY exists).
Note that the common SUSY scale, $M_{\rm{susy}}$,
fixes in our scenarios the general soft squark mass scale $m_{\tilde{q}}$ 
(see eqs. (\ref{deltadefa},\ref{deltadefb})) 
and the first diagonal element of the chargino mass matrix $M_2$
(see eq. (\ref{Xmat})). 

The parameters are chosen as follows:
for $M_{\rm{susy}}$
we choose the three values  
$M_{\rm{susy}}=300, 500, 1000 \,GeV$, while for $\tan \beta$ we use the
values $\tan \beta = 2, 10, 30, 50$. For the gluino mass, characterized
by $x=m^2_{\tilde{g}}/M^2_{\rm{susy}}$, we take
$x=0.3,0.5,1,2$. Unless otherwise stated, the $\mu$ parameter and the 
mass of the charged Higgs boson  $M_{H^-}$ are fixed to be
$\mu = 300 \,GeV$ and $M_{H^-} = 300 \,GeV$.

While in the first part of our analysis we  
set, following ref.\cite{GGMS}, all diagonal soft entries in 
$m^2_{\,d,\,LL}$, $m^2_{\,d,\,RR}$, and $m^2_{\,u,\,RR}$
equal to the common soft squark mass scale $m^2_{\tilde{q}}$,
we relax this condition in the second part of our analysis. 

We point out that the present bound on the mass of the lightest neutral
Higgs boson
requires a non-vanishing 
mixing $(m^2_{u,LR})_{33}$ among the stop-squarks.
For our choices of the parameters, the MSSM bound coincides with
the bound on the SM Higgs boson.

We note that there are two contributions to the stop-squark mixing, namely
the `soft' contribution $(m^2_{u,LR})_{33}$ and the $F$-term
contribution  ($-\mu \, m_t \, \cot \beta$).  In a general unconstrained 
MSSM, 
the soft contribution
does not scale with $m_t$. However, following common notation, 
we parametrize  the stop mixing term in 
${\cal M}_u^2$ (see eq. (\ref{massmatrixu})) as
\begin{equation}
X_t \, m_t = (m^2_{u,LR})_{33}\, -\mu m_t \cot \beta \, .
\end{equation}
We fix the stop mixing parameter $X_t$ such that the mass of the lightest 
Higgs boson is at least $115 \, GeV$ to assure that the present Higgs bound 
is fulfilled in our analysis. We  use the program FeynHiggsFast 
\cite{heinemeyer} which determines the Higgs boson mass approximately
taking into account the complete one- and two-loop QCD 
corrections, the effects of the running top mass and of the Yukawa term
for the light Higgs boson. The input parameters of the FeynHiggsFast program,
are $\tan \beta$, 
the diagonal entries 
of the stop and the sbotton squark mass matrix, $M^2_{\tilde{t}_L}$,
$M^2_{\tilde{t}_R}$, $M^2_{\tilde{b}_L}$, and $M^2_{\tilde{b}_R}$,
the stop and the sbotton mixing parameters, $X_t$ and $X_b$,
the top mass $m_t$, the parameter $\mu$ and the charged Higgs boson
mass. 
For $\tan \beta = 10$ we choose  the following values for $X_t$
in dependence of the parameter $M_{{\rm susy}}$: 
$(M_{\rm{susy}},X_t) = (300 \,GeV, 470 \,GeV)$; $(500 \,GeV, 750 \,GeV)$; 
$(1000 \,GeV, 1200 \,GeV)$. 
(For $\mu = 300 \,GeV$ and $X_b =0$ we find Higgs boson
masses of $115.2 \,GeV$, $119.9 \,GeV$ and $121.1 \,GeV$, respectively.) 
The dependence of the Higgs boson mass on the parameters 
$\mu$ or $X_b$ is rather small within 
our parameter scenarios. 
Also for the $\tan \beta = 30$ and $50$ scenarios
we use the same values of $X_t$ for our convenience. In these cases
the chosen $X_t$ values imply slightly higher Higgs boson masses. 
We also note that within our choice of parameters,
the low $\tan \beta$ 
scenario, with $\tan \beta = 2$, already gets excluded by the bound
on the Higgs boson mass.
\subsection{First part of analysis}
In this part of the analysis only {\it one} off-diagonal entry
in the soft part of the squark mass matrices is different from zero.
We further  assume (as in ref.~\cite{GGMS}) 
that  all diagonal soft entries in 
$m^2_{\,d,\,LL}$, $m^2_{\,d,\,RR}$, and $m^2_{\,u,\,RR}$
are set to be equal to the common value $m_{\tilde{q}}^2=M^2_{\rm{susy}}$. 
Then we normalize the off-diagonal elements to $m_{\tilde{q}}^2$,
\begin{equation} 
\delta_{f,LL,ij} = \frac{(m^2_{\,f,\,LL})_{ij}}{m^2_{\tilde{q}}}\,, 
\hspace{1.0truecm}
\delta_{f,RR,ij} = \frac{(m^2_{\,f,\,RR})_{ij}}{m^2_{\tilde{q}}}\,, 
\hspace{1.0truecm}
(i \ne j) 
\label{deltadefa}
\end{equation}
\begin{equation} 
\delta_{f,LR,ij} = \frac{(m^2_{\,f,\,LR})_{ij}}{m^2_{\tilde{q}}}\,,
\hspace{1.0truecm}
\delta_{f,RL,ij} = \frac{(m^2_{\,f,\,LR})^\dagger_{ij}}{m^2_{\tilde{q}}}\,.
\phantom{\hspace{1.0truecm}
(i \ne j)} 
\label{deltadefb}
\end{equation}
We recall that the matrix $m_{u,LL}$ cannot be specified independently;
$SU(2)_L$ gauge invariance implies that 
$m_{u,LL} =  K m_{d,LL} K^\dagger$, where $K$ is the CKM
matrix. We also note that our $\delta$-quantities only include the
soft parts of the  matrix elements of the squark mass matrices,
while in ref. \cite{GGMS} also the $F$-term contributions are included
in the definition of the $\delta$-quantities.
\begin{figure}[t]
    \begin{center}
    \leavevmode
    \includegraphics[height=8cm,bb=100 366 382 704]{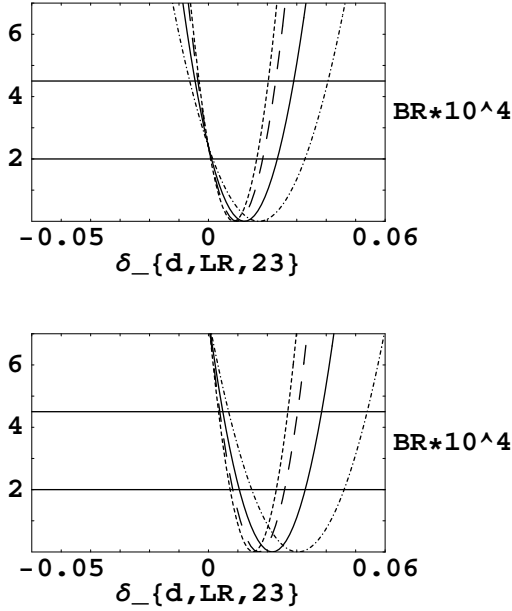}
    \vspace{2ex}
    \caption[f1]{       Dependence of $BR(B\rightarrow X_s\gamma)$ 
                        on $\delta_{d,LR,23}$. 
                        In the upper frame, only SM and gluino contributions
                        are considered. 
                        In the lower frame, the additional contributions 
                        from chargino, charged Higgs boson and neutralino are 
                        included.
                        The horizontal lines denote the experimental limits.
                        The different lines correspond                        
                        to different values of
                        $x=m^2_{\tilde{g}}/m^2_{\tilde{q}}$:
                        0.3 (short-dashed line), 
                        0.5 (long-dashed line), 1 (solid line), 
                        and 2 (dot-dashed line). The other parameters are
                        $\mu=300 \,GeV$, $\tan\beta=10$, $M_{H^-}=300 \,GeV$, 
                        $M_{\rm{susy}}=500 \,GeV$.}
    \label{fig:adI}
    \end{center}
\end{figure}
\begin{figure}[t]
    \begin{center}
    \leavevmode
    \includegraphics[height=8cm,bb=100 366 382 704]{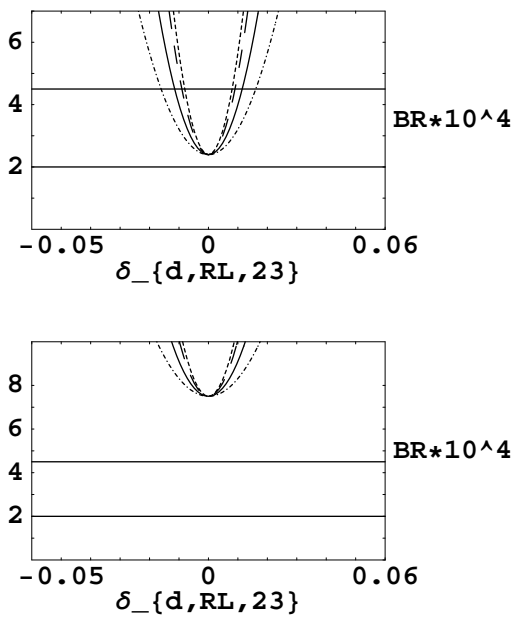}
    \vspace{2ex}
    \caption[f1]{Same as in fig. \ref{fig:adI} when $\delta_{d,RL,23}$
                        is the only non-zero off-diagonal squark mass entry.} 
    \label{fig:adII}
    \end{center}
\end{figure}
In figs. 
\ref{fig:adI} and  \ref{fig:adII},
we show the dependence of the branching ratio of 
$B \rightarrow X_s\,\gamma$
on the flavour-violating parameters 
$\delta_{d,LR,23}$ and $\delta_{d,RL,23}$, respectively. 
The upper frame in each figure
is borrowed from \cite{BGHW}, i.e. we consider only SM  and 
gluino contributions. As $\delta_{d,LR,23}$ and $\delta_{d,RL,23}$
generate the five-dimensional dipole operators
${\cal O}_{7\tilde{g},\tilde{g}}$ and 
${\cal O}'_{7\tilde{g},\tilde{g}}$, it is not surprising that they
get stringently bounded. 
We should note that at this level of the analysis
there is no dependence of these bounds on $\mu$ or $\tan \beta$. 
Such a dependence could result from the term
$(F_{d,LR})_{33}$, but only when $\delta_{d,LL,23}$ and $\delta_{d,RR,23}$
are turned on. We will discuss this point in more detail 
at the end of this section and
in the second part of our analysis.
In the lower frame of figs. \ref{fig:adI} and  \ref{fig:adII},
we also include the contributions from
charginos, charged Higgs bosons and neutralinos. Comparing the 
branching ratio in the two frames at  
$\delta_{d,LR,23}=0$ and $\delta_{d,RL,23}=0$ (which corresponds to
switching off the gluino contribution), one concludes that the 
combined contribution from charginos, neutralinos and charged Higgs bosons
is of the same order as the SM contribution. 
A detailed investigation shows that the neutralino
contribution
is negligible, while the contributions from the chargino and charged 
Higgs boson 
are similar in magnitude; both interfere constructively with the SM
contributions for the specific choice of parameters.
However, as the gluino yields, intrinsically, the dominant contribution by far, 
the bounds
$\delta_{d,LR,23}$ and $\delta_{d,RL,23}$ are only marginally modified
by chargino, neutralino and charged Higgs boson contributions.    
A comment concerning the different shapes of the curves in figs.
\ref{fig:adI} and  \ref{fig:adII} is in order.
In fig. \ref{fig:adII}, with 
non-vanishing $\delta_{d,RL,23}$, the gluino contribution is induced by the  
primed-type operator ${\cal O}'_{7\tilde{g},\tilde{g}}$ and therefore
does not interfere with the contributions from the other particles,
as these induce unprimed operators in the first place.
In contrast, in fig. \ref{fig:adI}, which shows the case of 
non-zero $\delta_{d,LR,23}$, the gluino contribution is of the unprimed type 
and
therefore interferes with the other contributions.

We also tried to derive analogous bounds on 
$\delta_{d,LL,23}$, $\delta_{d,RR,23}$, 
$\delta_{u,LR,23}$, $\delta_{u,RL,23}$,
$\delta_{u,RR,23}$ and also
on $\delta_{u,LR,33}$ and $\delta_{u,LR,22}$. 
In the chargino sector the latter 
diagonal elements, together with the usual CKM mechanism, also can induce 
flavour violation. 
The parameters of the up-squark mass matrix give rise to chargino 
contributions that lead  only to dimension six dipole operators,
which inherently are not very large. 
For our choices of $\mu$, $M_{\rm{susy}}$ and $\tan \beta$,
this was confirmed numerically. Therefore, no stringent bounds
are obtained for the soft parameters in the up-squark mass 
matrix\footnote{In \cite{newpaper} the authors derived a rather stringent  
bound on a quantity proportional to $\delta_{u,LR,33}$ in the case of a 
small chargino mass of $100 \,GeV$. However, they include the small
CKM factor $K_{ts}^* K_{tb} \approx 1/30$ in the definition of their
quantity.}. 
The remaining parameters of the down-squark mass matrix, i.e.  
$\delta_{d,LL,23}$ and $\delta_{d,RR,23}$, play an interesting role.
They not only generate  contributions to the six-dimensional operators 
in (\ref{gmagnopb}), but,
together with the chirality changing term $(F_{d,LR})_{33}$, they also
induce contributions to the five-dimensional gluino operators
in (\ref{gmagnopg}). For the values of $\mu$ and $\tan \beta$ used in our analysis, 
the coefficients of the five-dimensional 
operators turn out to be rather small. Thus, no stringent bounds on 
$\delta_{d,LL,23}$ and $\delta_{d,RR,23}$ are obtained. 

Summarizing the first part of our analysis, we conclude 
that $\delta_{d,LR,23}$ and $\delta_{d,RL,23}$
are the only parameters that get significantly constrained 
by the measurement of the branching ratio of $B \rightarrow X_s \gamma$.
\subsection{Second part of analysis}
We now explore the problem of whether the separate bounds on 
$\delta_{d,LR,23}$ and $\delta_{d,RL,23}$, obtained in the first part,
remain stable if the various soft parameters are varied simultaneously.
The analysis is based on the assumption that the soft terms in the
squark mass matrices have the hierarchical structure that the diagonal entries
in $m^2_{\,d,\,LL}$, $m^2_{\,d,\,RR}$, and $m^2_{\,u,\,RR}$ are larger
than the off-diagonal matrix elements (including  $m^2_{\,d,\,LR}$ 
and $m^2_{\,u,\,LR}$). In contrast to the first part of the 
analysis, we will allow for a non-degeneracy of the diagonal elements in the
matrices $m^2_{\,d,\,LL}$, $m^2_{\,d,\,RR}$, and $m^2_{\,u,\,RR}$.
To implement this, we define 
$\delta$-quantities in addition to those in eqs. 
(\ref{deltadefa}) and (\ref{deltadefb}), which
parametrize this non-degeneracy:
\begin{equation} 
\delta_{f,LL,ii} = 
\frac{(m^2_{\,f,\,LL})_{ii} - m^2_{\tilde{q}}}{m^2_{\tilde{q}}}\,, 
\hspace{1.0truecm}
\delta_{f,RR,ii} = 
\frac{(m^2_{\,f,\,RR})_{ii} - m^2_{\tilde{q}}}{m^2_{\tilde{q}}}\,, 
\label{deltadefc}
\end{equation}
Unless otherwise stated, 
the diagonal $\delta$-parameters (in eq. (\ref{deltadefc})) are varied
in the interval $[-0.2,0.2]$. On the other hand, the
off-diagonal ones
(in eqs. (\ref{deltadefa}) and (\ref{deltadefb})) are varied
in the interval $[-0.5,0.5]$, by use of a Monte Carlo program. There are,
however, two exceptions. First,  we do not vary  
those off-diagonal $\delta$'s with an index $1$; the latter $\delta$'s
we set to zero, since they are severely constrained by kaon decays
(see for example \cite{GGMS}). 
Second,
as mentioned earlier, also $(m^2_{u,LR})_{33}$ is not varied, but fixed
such that the mass of the lightest neutral Higgs boson gets heavy enough to be
compatible with experimental bounds.

In our Monte Carlo analysis we  plot those events, 
corresponding to  $2.0\times 10^{-4}\leq BR(B \rightarrow X_s \gamma) 
\leq 4.5\times 10^{-4}$, which is the range allowed
by the CLEO measurement. Note that we do not include
recent preliminary data \cite{BSGMEASURE} in our analysis. 
Furthermore, we have 
made sure that our events correspond to squark masses that are real 
and lie above 
$150 \,GeV$. 
The dependence of the bounds on this specific 
choice is  discussed below.
\begin{figure}[t]
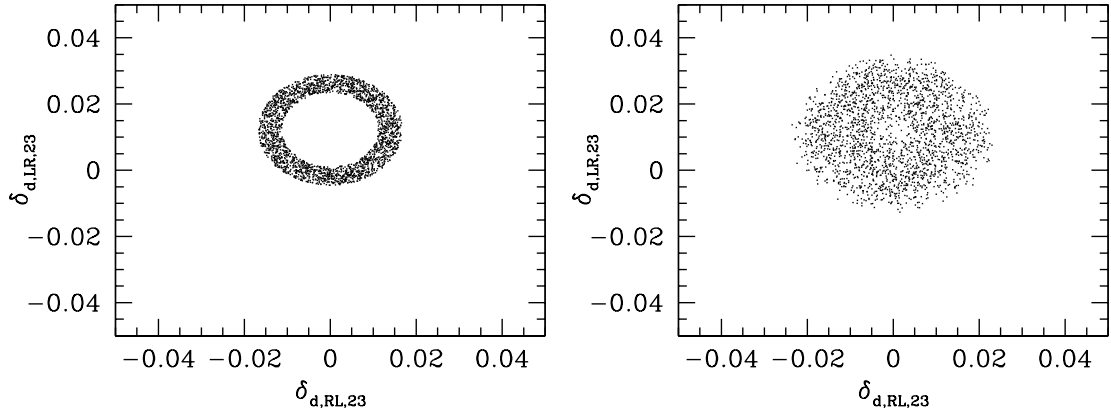

    \begin{center}
    \leavevmode
    \includegraphics[height=6cm,bb= 20 145 560 586]{bild3oben.epsi}
    \includegraphics[height=6cm,bb= 20 145 560 586]{bild3unten.epsi}
    \vspace{2ex}
    \caption[f1]{Contours in the $\delta_{d,LR,23}$ - $\delta_{d,RL,23}$ plane.
             In the left frame, $\delta_{d,LR,23}$ and 
             $\delta_{d,RL,23}$ are the
             only flavour-violating parameters. In the right frame, 
             we allow also for 
             non-vanishing $\delta_{d,LL,23}$ and $\delta_{d,RR,23}$. We only
             consider SM and gluino contributions and the other 
             parameters are $\mu=300\,GeV$, $\tan\beta=10$, 
             $M_{\rm{susy}}=500\,GeV$ and $x=1$.}
    \label{fig:ad23ad32}
    \end{center}
\end{figure}
We start with the following parameter set:  $\mu=300\,GeV$,
$M_{H^-}=300\, GeV$, $\tan\beta=10$,
$M_{\rm{susy}}=500\,GeV$, 
$x= m_{\tilde{g}}^2 \, / \, M_{\rm{susy}}^2 = 1$ and
$X_t=750\,GeV$. 
In fig. \ref{fig:ad23ad32}, we only consider SM and gluino 
contributions. In the left frame  we  present the constraints  on
$\delta_{d,LR,23}$ and $\delta_{d,RL,23}$ when these are 
the only flavour-violating soft parameters;
the diagonal $\delta$-parameters defined in eq. (\ref{deltadefc}) 
are also switched off.
As expected from the first part
of our analysis, stringent constraints are obtained.
The hole inside the dotted area represents values of
$\delta_{d,LR,23}$ and $\delta_{d,RL,23}$
for which the branching ratio is too small to be compatible with
the measurements.
In the right frame we investigate interference effects that  arise when
$\delta_{d,LL,23}$ and $\delta_{d,RR,23}$ are switched on in addition to 
$\delta_{d,LR,23}$, $\delta_{d,RL,23}$. 
All of them are varied between $\pm 0.5$.
From fig. \ref{fig:ad23ad32} we find that the bounds on 
$\delta_{d,LR,23}$ and $\delta_{d,RL,23}$ cannot be softened 
significantly by non-zero values of 
$\delta_{d,LL,23}$ and $\delta_{d,RR,23}$, although
these $\delta$-parameters, which individually give  rise to six-dimensional
operators, generate five-dimensional operators through the interplay
with the $F$-term $(F_{d,LR})_{33}$.
As already discussed in the first part of the analysis, for moderate values
of $\mu$ and $\tan \beta$, the contribution to the Wilson coefficient of the 
five-dimensional operator is rather small. 
\begin{figure}[t]
    \begin{center}
    \leavevmode
    \includegraphics[height=6cm,bb=20 145 560 586]{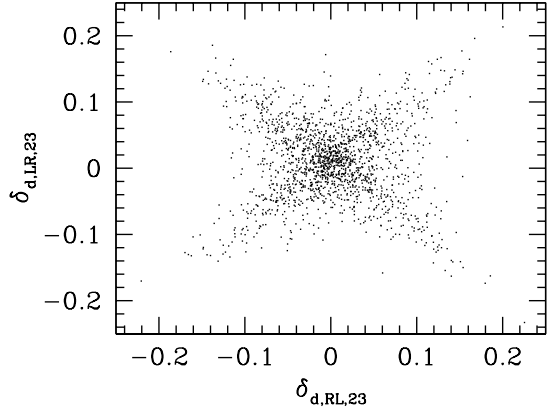}
    \vspace{2ex}
    \caption[f1]{Contours in the $\delta_{d,LR,23}$ - $\delta_{d,RL,23}$ 
           plane, where
           $\delta_{d,LR,23}$, $\delta_{d,RL,23}$, $\delta_{d,LL,23}$ and 
           $\delta_{d,RR,23}$ are the only flavour-violating parameters;
           $\delta_{d,LR,22}$, $\delta_{d,LR,33}$ are also non-vanishing. 
           We only include SM and gluino contributions; the other 
             parameters are $\mu=300\,GeV$, $\tan\beta=10$, 
             $M_{\rm{susy}}=500\,GeV$ and $x=1$.} 
    \label{fig:ad23ad32q23d23ad22ad33}
    \end{center}
\end{figure}

\begin{figure}[t]
    \begin{center}
    \leavevmode
    \includegraphics[height=6cm,bb=20 145 560 586]{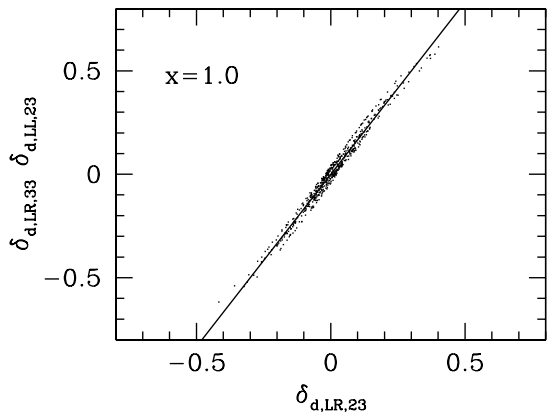}
    \vspace{2ex}
    \caption[f1]{The parameters $\delta_{d,LR,23}$ and 
              $\delta_{d,LR,33} \cdot \delta_{d,LL,23}$, 
              which are compatible
              with the data on $B \to X_s \gamma$ are shown by dots.
              Values lying on the solid line lead to a vanishing contribution
              of the five-dimensional operator 
              ${\cal O}_{7\tilde{g},\tilde{g}}$  
              in the MIA. See text. We only switch on SM and gluino
              contributions;
              the other parameters are $\mu=300\,GeV$, $\tan\beta=10$, 
              $M_{\rm{susy}}=500\,GeV$ and $x=1$.} 
    \label{fig:linie}
    \end{center}
\end{figure}

The full power of the interference effects from different sources
of flavour violation is depicted in fig. 
\ref{fig:ad23ad32q23d23ad22ad33}, where we allow not only for non-zero 
 $\delta_{d,LR,23}$, $\delta_{d,RL,23}$, $\delta_{d,LL,23}$ and 
$\delta_{d,RR,23}$ but also
for non-vanishing $\delta_{d,LR,22}$, $\delta_{d,LR,33}$. 
All these parameters are varied between
$\pm 0.5$. As can be seen, the bounds on $\delta_{d,LR,23}$ and 
$\delta_{d,RL,23}$ get destroyed dramatically. The reason is that
there are now new contributions to the five-dimensional dipole
operators. As an example,   
the combined effect of $\delta_{d,LR,33}$ and
$\delta_{d,LL,23}$ leads  to a contribution to the Wilson coefficient
of the operator ${\cal O}_{7\tilde{g},\tilde{g}}$. The sign of
this contribution can be different from the one generated by
$\delta_{d,LR,23}$. As a consequence, the bound on $\delta_{d,LR,23}$
gets weakened. To illustrate this more quantitatively, we assume
for the moment that   there are only these
two sources that  can generate  ${\cal O}_{7\tilde{g},\tilde{g}}$,
i.e.  we switch off the other $\delta$-quantities.
If $\delta_{d,LR,23}$ is larger than the individual bound from the
first part of the analysis, it is necessary that the product
of  $\delta_{d,LR,33}$ and $\delta_{d,LL,23}$ is also relatively
large; only in this case can the two sources lead to a branching
ratio compatible with experiment. This feature is illustrated 
in fig. \ref{fig:linie}; only values of  $\delta_{d,LR,23}$
and values of  $\delta_{d,LR,33} \cdot \delta_{d,LL,23}$ which are 
strongly correlated lead to an acceptable branching ratio. 
As clearly visible from fig. \ref{fig:linie}, the correlation
between the two sources for  ${\cal O}_{7\tilde{g},\tilde{g}}$, 
is essentially linear. This implies that the linear combination
\begin{equation} 
 \delta_{d,LR,23} + f  \delta_{d,LR,33} \cdot \delta_{d,LL,23}
\label{combi}
\end{equation} 
gets constrained, if $f$ is chosen appropriately. Stated differently, 
the Wilson coefficient of the operator  ${\cal O}_{7\tilde{g},\tilde{g}}$
is essentially proportional to the combination (\ref{combi}).
This implies in turn that for the values of the parameters we are using 
at the moment ($\mu=300\,GeV$,
$M_{H^-}=300\,GeV$, $\tan\beta=10$
$M_{\rm{susy}}=500\,GeV$, 
$x= m_{\tilde{g}}^2 \, / \, M_{\rm{susy}}^2 = 1$,
$X_t=750\,GeV$),
the Wilson coefficient is well approximated by its double 
mass insertion expression. The coefficient $f$, which can be
read off from this expression, 
depends on the parameter $x= m_{\tilde{g}}^2/M^2_{\rm{susy}}$ and
reads
\begin{equation}
f(x)=     \frac{1 + 9 x - 9 x^2 - x^3 +6x(1+x)\log x}{
           (1-x)[5x^2-4x-1-2x(x+2)\log x]} \, .
\label{eq:f(x)}
\end{equation}
The numerical values of $f(x)$ for some values of $x$ read $0.74$ for $x=0.3$,
$0.68$ for $x=0.5$,  $0.60$ for $x=1.0$ and $0.52$ for $x=2.0$, respectively.

The solid line in fig.  \ref{fig:linie} represents 
pairs ($\delta_{d,LR,23}$, $\delta_{d,LR,33} \cdot \delta_{d,LL,23}$)
for which the combination in eq. (\ref{combi}) is zero. The points
scattered around this line therefore represent Monte Carlo events for which 
this combination is small.
We now turn back to the scenario of fig. \ref{fig:ad23ad32q23d23ad22ad33}
in which all the parameters  
$\delta_{d,LR,23}$, $\delta_{d,RL,23}$, $\delta_{d,LL,23}$,
$\delta_{d,RR,23}$, $\delta_{d,LR,22}$, $\delta_{d,LR,33}$
are varied simultaneously. In this case, the linear combinations

\begin{eqnarray}
LC_1 & = & \delta_{d,RL,23}+f(x)\delta_{d,RR,23}\cdot\delta_{d,RL,33} 
           +f(x)\delta_{d,RL,22}\cdot\delta_{d,LL,23},\nonumber\\ 
LC_2 & = & \delta_{d,LR,23}+f(x)\delta_{d,LR,22}\cdot\delta_{d,RR,23}
           +f(x)\delta_{d,LL,23}\cdot\delta_{d,LR,33}, 
\label{deflc1lc2}
\end{eqnarray}
are expected to get constrained. 

\begin{figure}[t]
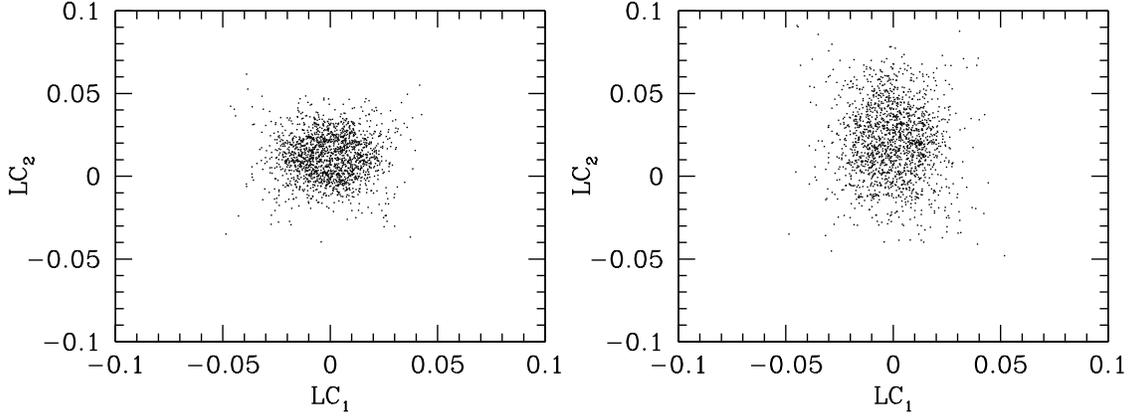

    \begin{center}
    \leavevmode
    \includegraphics[height=6cm,bb= 20 145 560 586]{bild5oben.epsi}
    \includegraphics[height=6cm,bb= 20 145 560 586]{bild5unten.epsi}
    \vspace{2ex}
    \caption[f1]{Contours in the $LC_1$-$LC_2$ plane with
           $\delta_{d,LR,23}$, $\delta_{d,RL,23}$, 
           $\delta_{d,LL,23}$, $\delta_{d,RR,23}$, 
           $\delta_{d,LR,22}$, $\delta_{d,LR,33}$,
           $\delta_{u,LR,23}$, 
           $\delta_{u,RL,23}$, 
           $\delta_{u,LL,23}$, 
           $\delta_{u,RR,23}$, and
           $\delta_{u,LR,22}$
           all non-vanishing. In the left frame,
           we consider only SM and gluino contributions whereas in the 
           right frame
           we also include chargino, charged Higgs boson and 
           neutralino contributions.
           The values of the other parameters are 
           $\mu=300\,GeV$, $\tan\beta=10$, 
           $M_{H^-}=300\,GeV$, $M_{\rm{susy}}=500\,GeV$ and $x=1$.} 
    \label{fig:lc1lc2}
   \end{center} 
\end{figure}

In fig.  \ref{fig:lc1lc2} we show the allowed region for $LC_1$ and $LC_2$. 
There, we allow all 
non-diagonal $\delta$-parameters to vary between $\pm 0.5$. 
In addition, we also allow for non-equal diagonal soft entries, by varying
the parameters $\delta_{f,LL,ii}$ and $\delta_{f,RR,ii}$ between   
$\pm 0.2$. With the latter choice
we still guarantee the hierarchy between diagonal and off-diagonal entries,
but we get rid of the unnatural assumption of degenerate diagonal entries.
In the left frame, we 
include only SM and  gluino contributions. We find that 
the linear combinations $LC_1$ and $LC_2$ indeed get stringently bounded.
In the right frame of fig. \ref{fig:lc1lc2} we test the resistance of 
these bounds 
when the additional contributions (i.e., 
those from charginos, charged Higgs bosons and neutralinos) are turned 
on. In this case also 
$\delta_{u,LR,23}$, $\delta_{u,RL,23}$,  $\delta_{u,LL,23}$,
$\delta_{u,RR,23}$ and 
$\delta_{u,LR,22}$ are varied in the range $\pm 0.5$.
We find that the bound on $LC_1$ remains unchanged, while the one
on $LC_2$ gets somewhat weakened. This feature is expected, because
charginos and charged Higgs bosons contribute to
unprimed operators at first place. At this point we should stress
that these plots were obtained by choosing the renormalization
scale $\mu_b=4.8\,GeV$ and by requiring all squark masses to be
larger than $150 \,GeV$. We  checked that 
the bounds on $LC_1$ and
$LC_2$ remain practically unchanged when the renormalization
scale is varied between $2.4\, GeV$ and $9.6\, GeV$; they are also insensitive
to the value of the required minimal squark mass, as we found by 
changing 
$m_{\rm{squark\,min}}$ from $150\, GeV$ to $100\, GeV$ or $250\, GeV$. 
Moreover, we also checked 
whether the restriction to  $\mu=+300$ $GeV$ scenario is 
too severe: we redid the complete analysis 
for $\mu=-300\,GeV$ and confirmed that ther are  
no differences between the results of 
these two choices.

Two remarks are in order: 
First, one might wonder why we did not include 
terms like  
$\delta_{d,RR,33}\cdot\delta_{d,LR,23}$ in $LC_1$ and
$LC_2$, which would result into more complicated combinations. 
As we are allowing for nonequal diagonal soft entries, 
these terms give in principle 
additional contributions to the five dimensional operators.
However, as the diagonal $\delta$-parameters are only varied between
$\pm 0.2$, their influence on the Wilson coefficients is numerically small. 
For this reason, the simpler combinations $LC_1$ and $LC_2$,
defined in eqs. (\ref{deflc1lc2}), are sufficiently constrained
and we prefer to give bounds on these quantities.\\
Second, if we got rid of the
hierarchy of diagonal and off-diagonal entries in the squark mass matrices, 
stringent bounds on the simple combinations $LC_1$ and $LC_2$ 
certainly would no longer exist, simply because there would then be more 
contributions to the five-dimensional operators of similar 
magnitude.
In this case, however, the $full$ Wilson coefficients
of the five-dimensional operators  
still would be stringently constrained by the experimental
data on $B \to X_s \gamma$. Unfortunately, in this case not much information
can be extracted
for the individual soft parameters or simple combinations thereof.

Finally, we extend our analysis to other values of the input parameters.
So far, we found that the combinations
$LC_1$ and $LC_2$ (see eqs. (\ref{deflc1lc2}))
are stringently bounded in the scenario
characterized by the input values 
$\mu=300 \,GeV$, $M_{H^-}=300\,GeV$, $\tan\beta=10$,
$M_{\rm{susy}}=500\,GeV$, 
$x= m_{\tilde{g}}^2 \, / \, M_{\rm{susy}}^2 = 1$ and
$X_t=750\,GeV$. It is conceivable that the bounds on $LC_1$ and
$LC_2$ can get considerably  weakened in other scenarios.
Therefore, we analyse the bounds on the soft parameters within the following 
parameter sets: ($M_{\rm{susy}}, X_t$) $=$ $(300\,GeV, 470\, GeV)$,
$(500\, GeV, 750\, GeV)$, $(1000\, GeV, 1200\, GeV)$.
For $\tan\beta $ we explore the values: $\tan \beta =  10,\,  30,\,  50.$
Furthermore, the gluino mass $m_{\tilde{g}}$ is varied over the values
$ x = m_{\tilde{g}}^2 \, / \, M_{\rm{susy}}^2 = 0.3\,, \, 0.5\,,\,  1\,,\, 2$. 
\begin{figure}[t]
    \begin{center}
    \leavevmode
    \includegraphics[height=4.2cm,bb= 20 145 560 586]{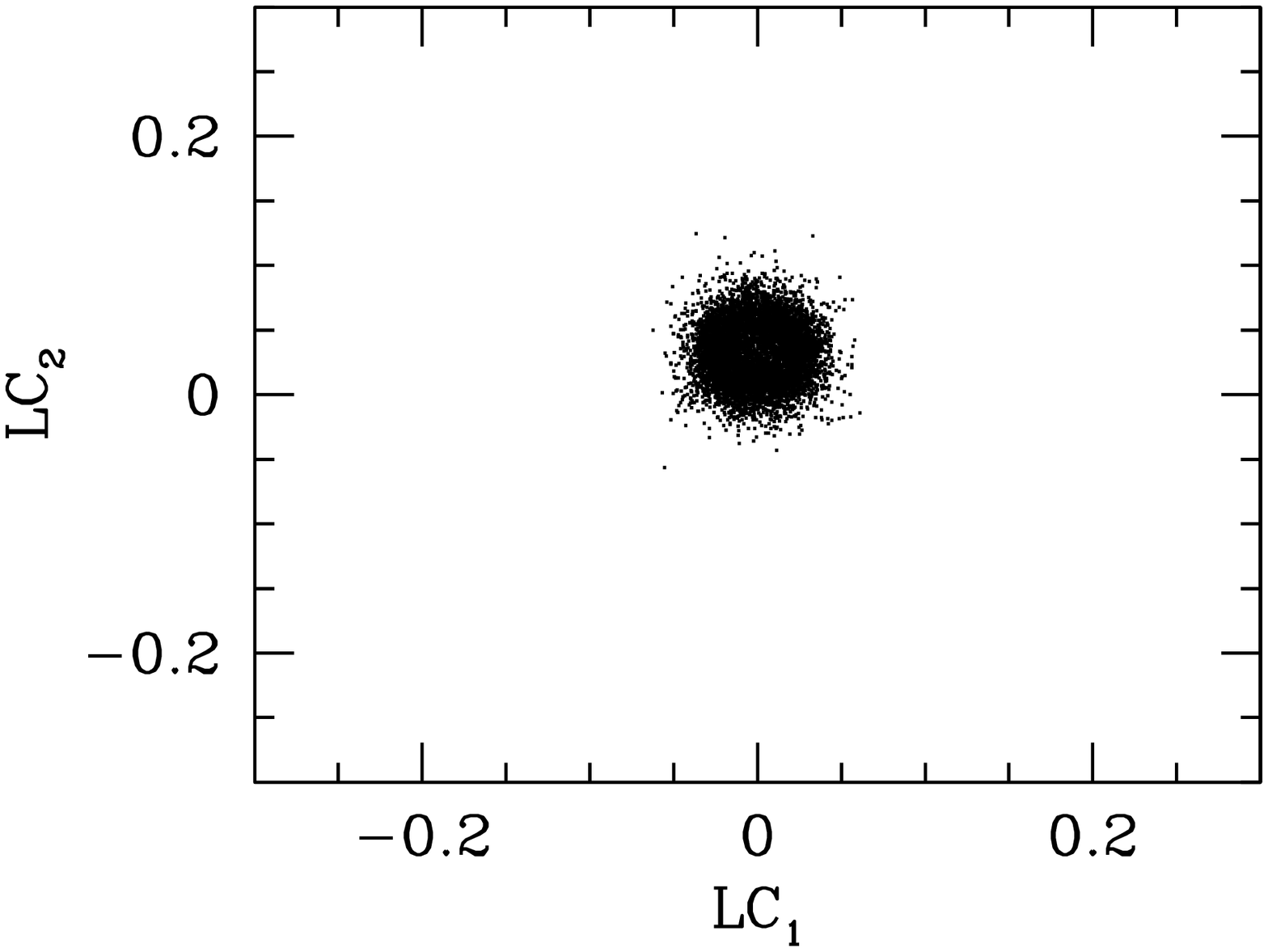}
    \includegraphics[height=4.2cm,bb= 20 145 560 586]{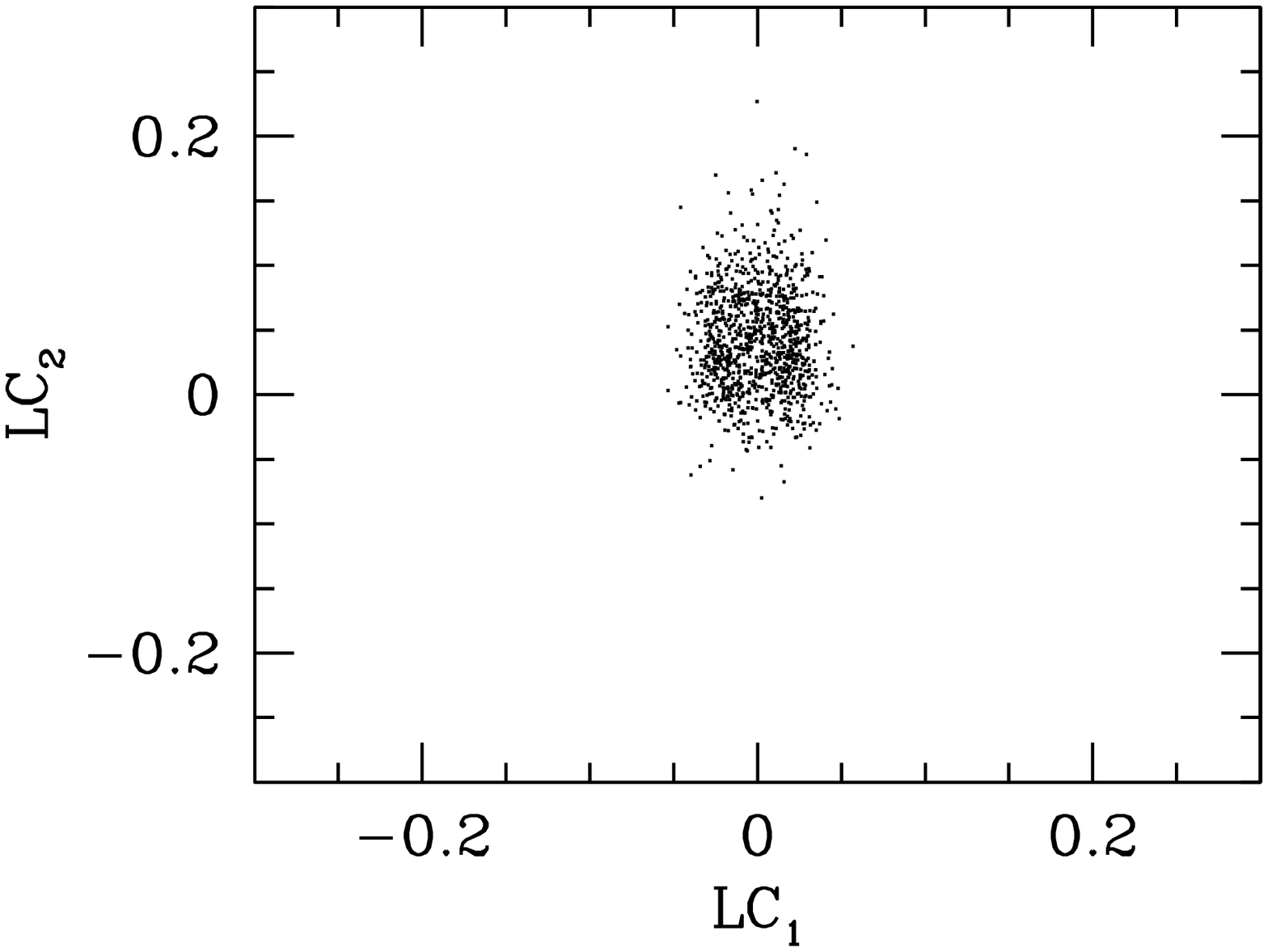}
    \includegraphics[height=4.2cm,bb= 20 145 560 586]{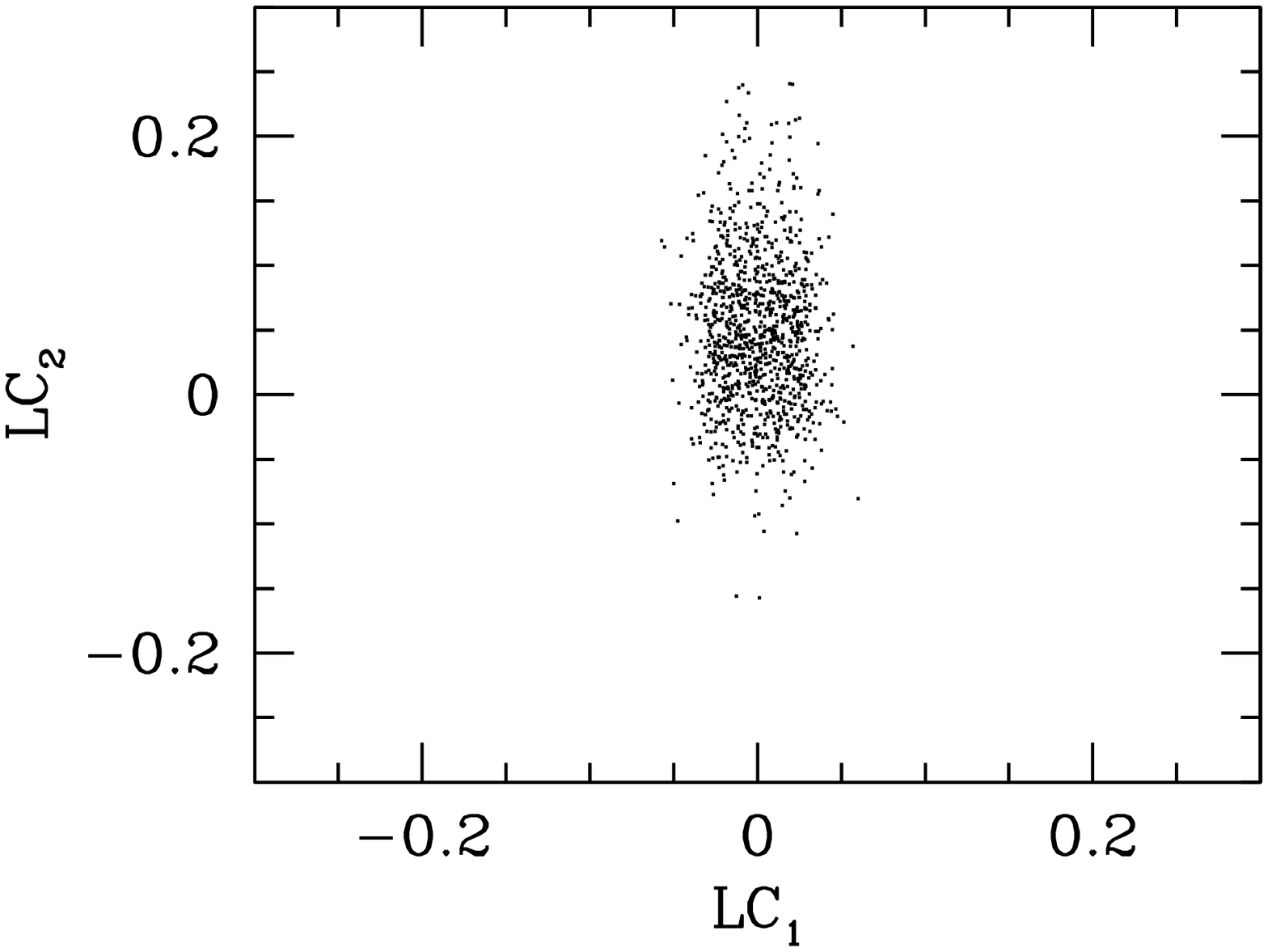}
    \includegraphics[height=4.2cm,bb= 20 145 560 586]{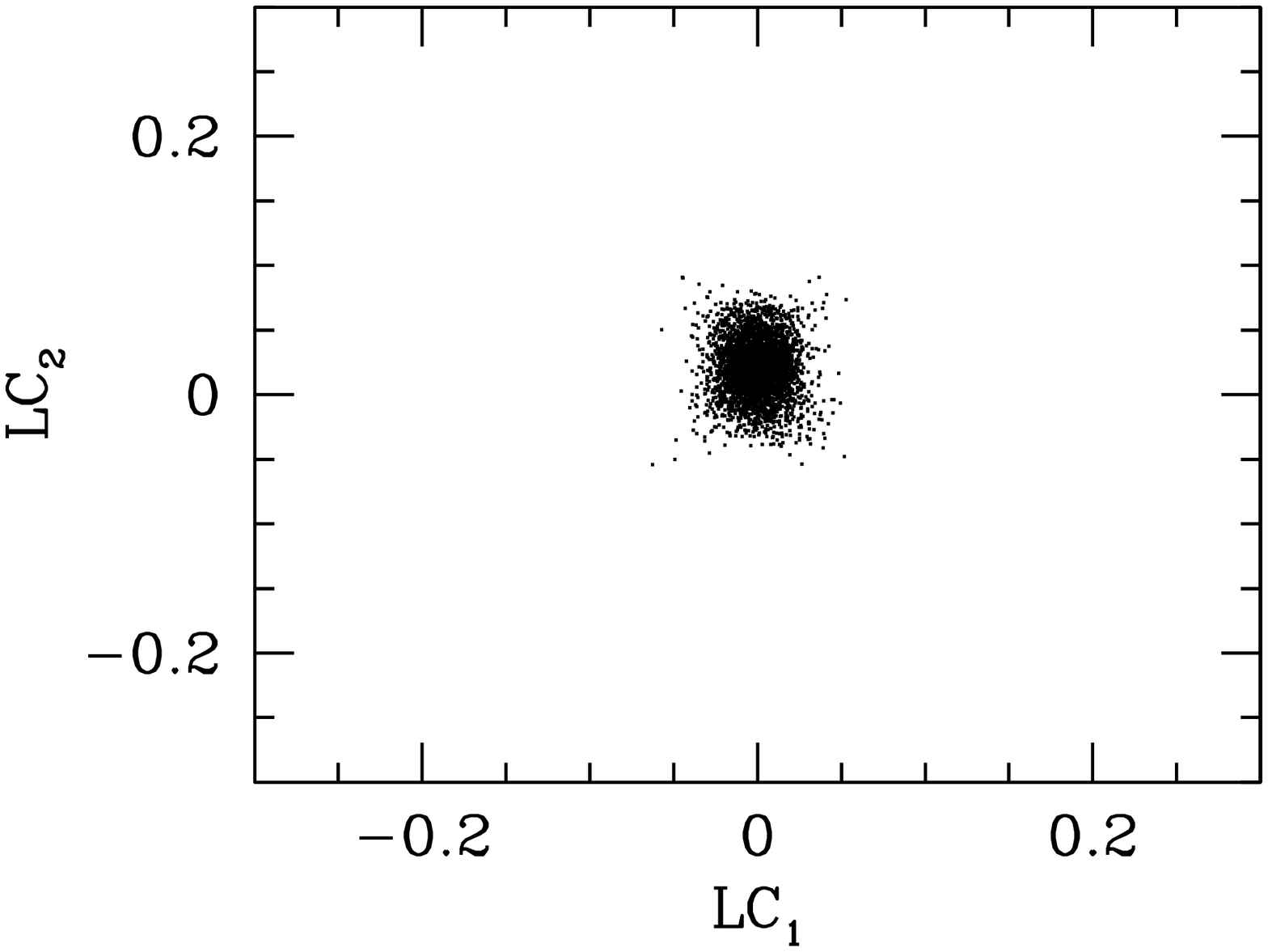}
    \includegraphics[height=4.2cm,bb= 20 145 560 586]{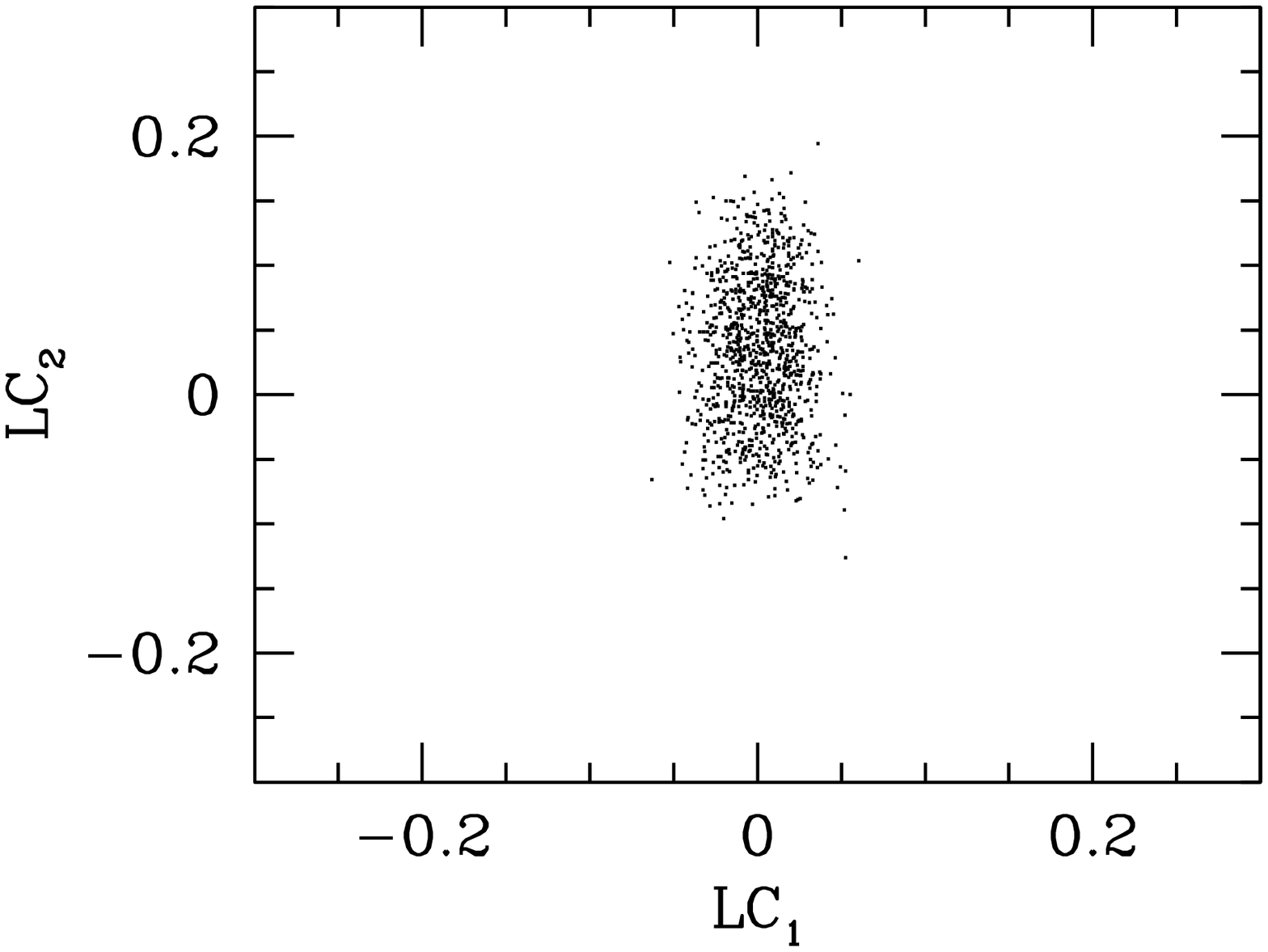}
    \includegraphics[height=4.2cm,bb= 20 145 560 586]{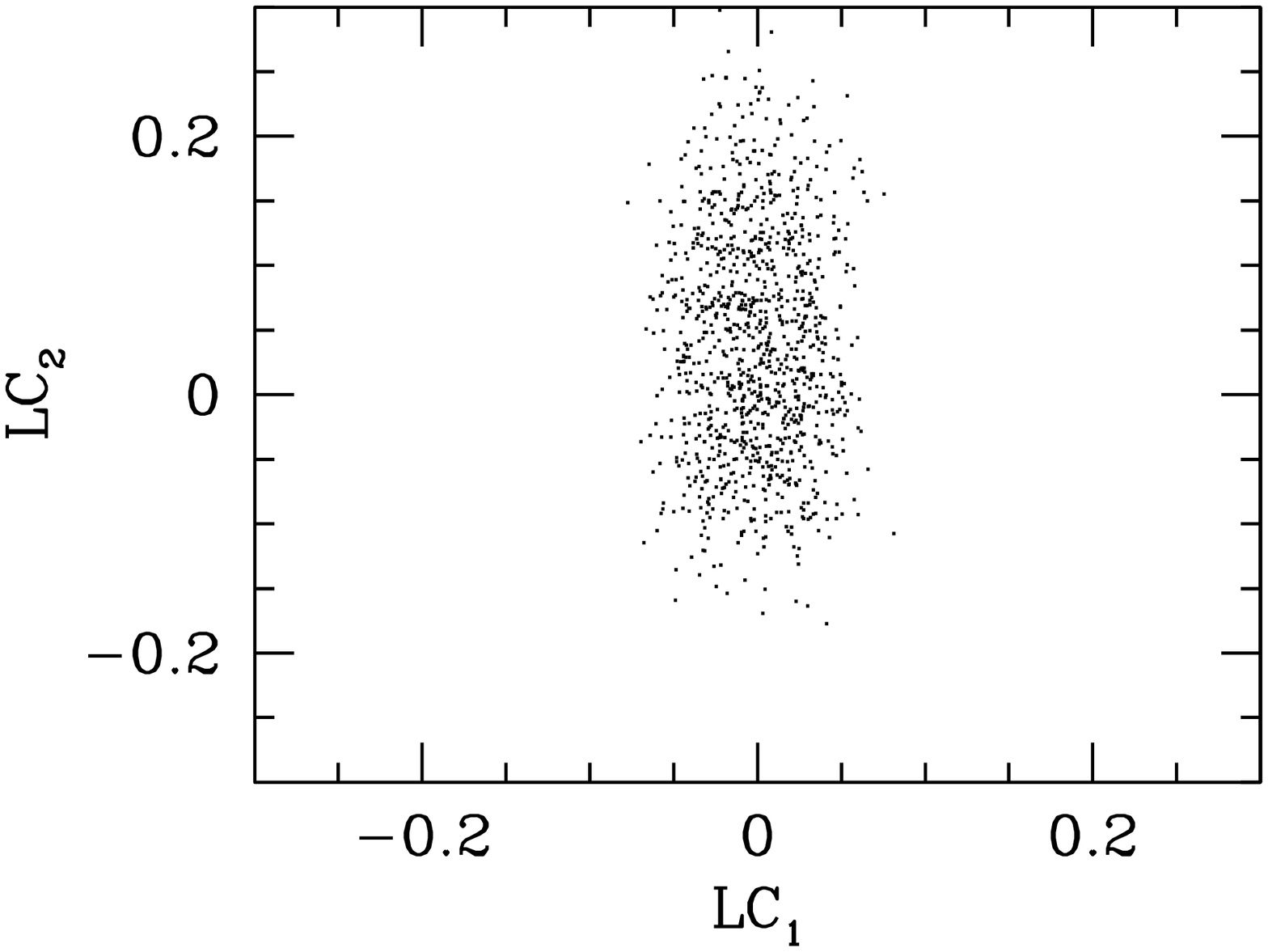}
    \includegraphics[height=4.2cm,bb= 20 145 560 586]{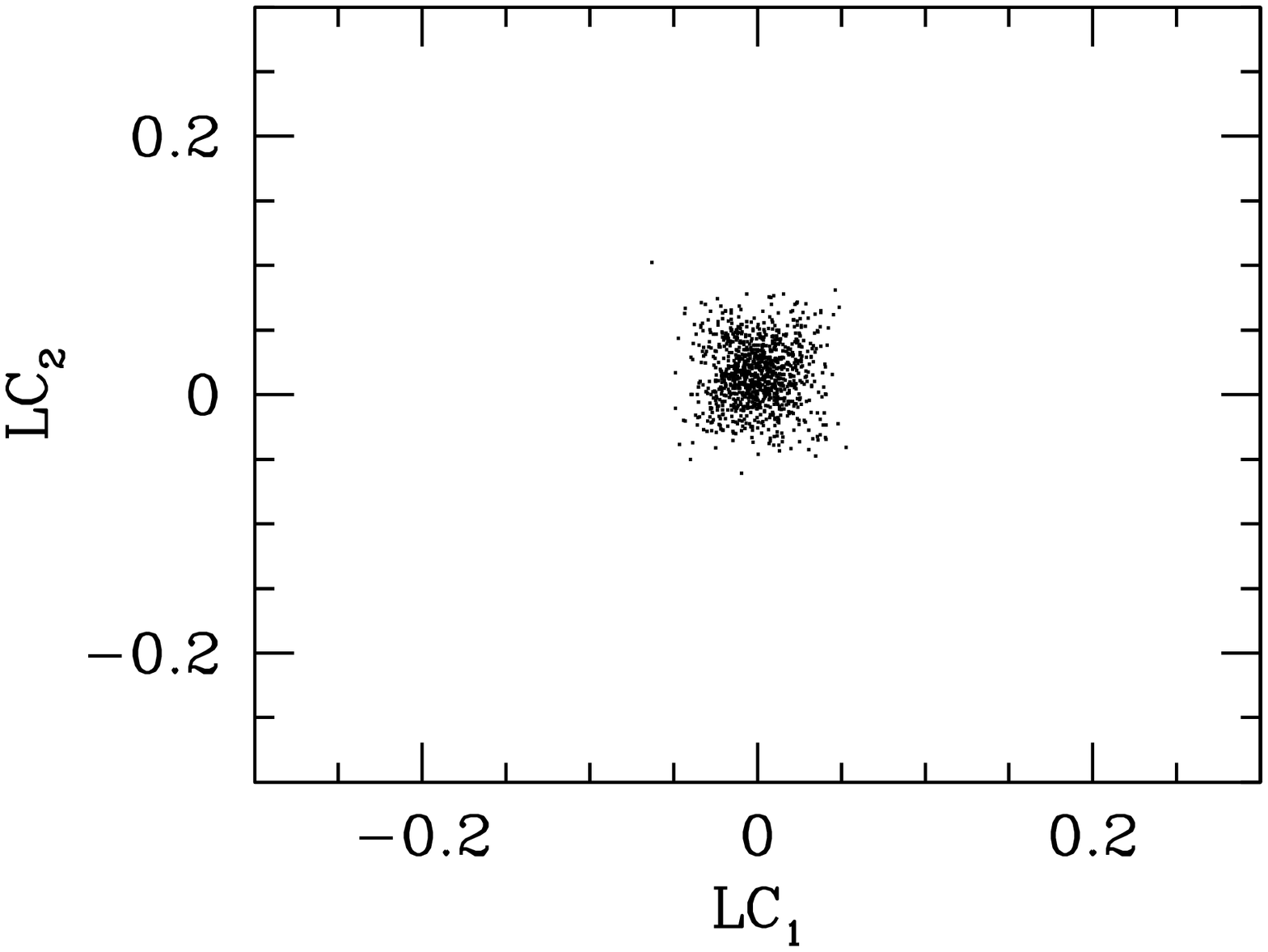}
    \includegraphics[height=4.2cm,bb= 20 145 560 586]{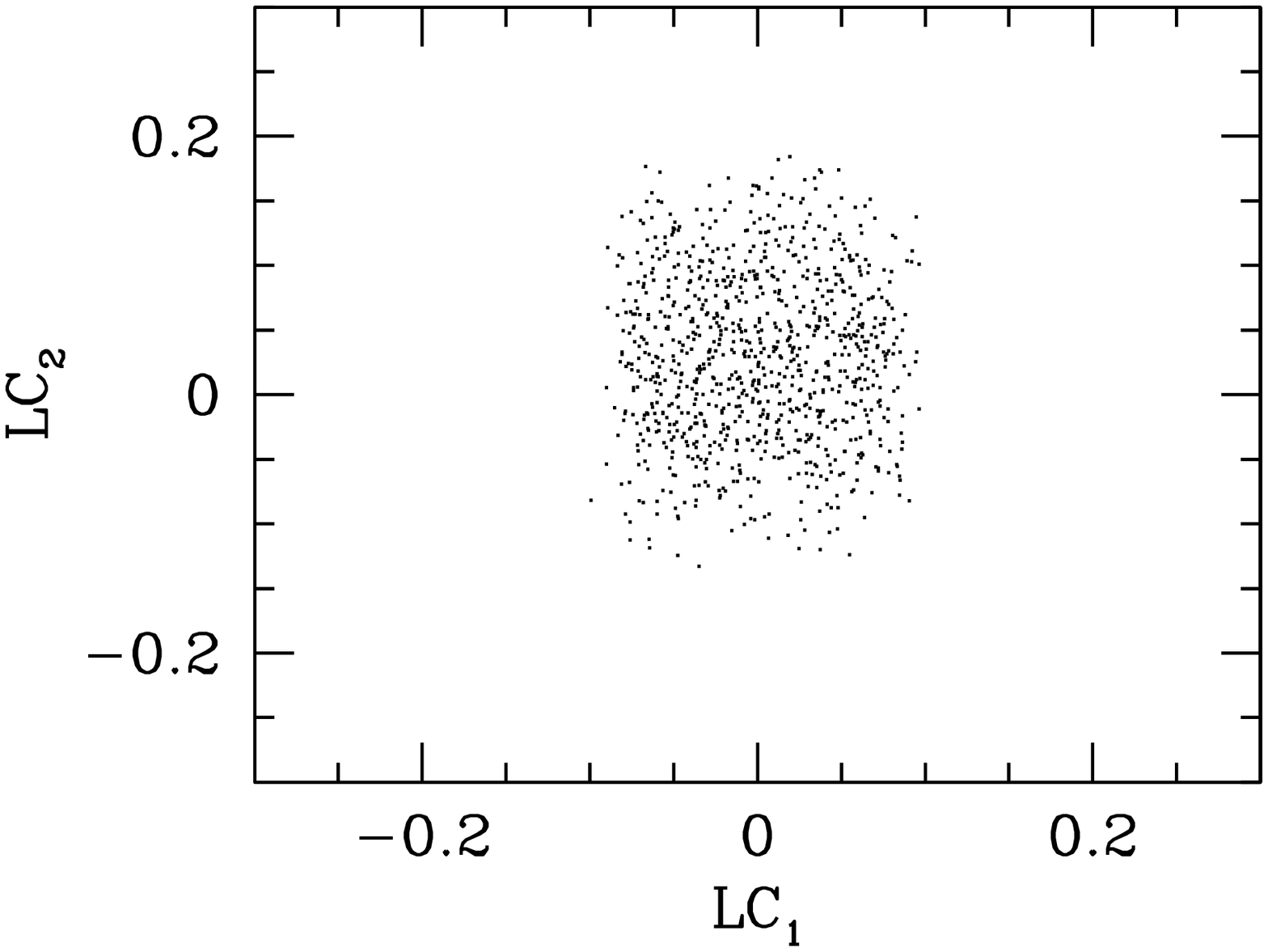}
    \includegraphics[height=4.2cm,bb= 20 145 560 586]{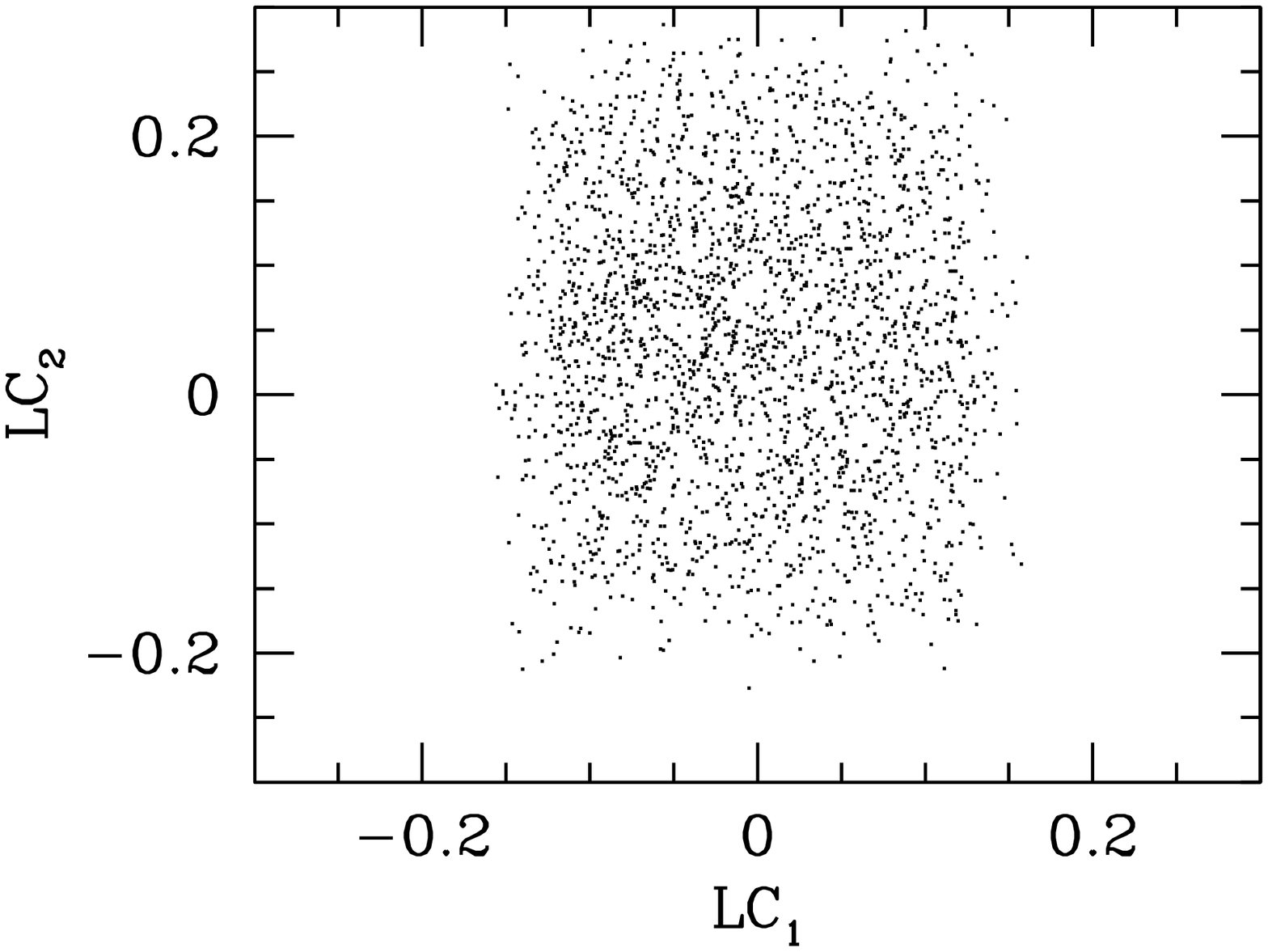}
   \vspace{2ex}
    \caption[f1]{$LC_1$ versus $LC_2$ in the $x=1$ scenario. 
             Vertically we vary the parameter $M_{{\rm susy}}$
             from $M_{{\rm susy}}=1000\, GeV$\, in the first line,
             via $M_{{\rm susy}}=500\, GeV$\, in the second   
             to $M_{{\rm susy}}=300\, GeV$\, in the last. 
             Horizontally we vary $\tan \beta$ from $10$ via $30$
             to $50$. The values of the other parameters are 
             $\mu=300\,GeV$ and $M_{H^-}=300\,GeV$.
             All contributions are switched on.}
    \label{allx1}
   \end{center} 
\end{figure}

Surprisingly, the constraints on $LC_1$ and $LC_2$ are completely stable 
over large parts of the parameter space. Within the 
$\tan \beta = 10$ 
scenario the bounds are essentially unchanged if the other two parameters 
$M_{{\rm susy}}$ and $x$, are varied over the complete range of values 
given above. For example, the independence from the
parameter $M_{{\rm susy}}$ within this scenario can be 
read off from the comparison of frames in the first
vertical line in fig. \ref{allx1}. 

However, fig. \ref{allx1} also illustrates that the bounds get 
significantly
weakened or even lost when $\tan \beta$ values as large as 
$30$ (second vertical line)
or $50$ (third vertical line) are chosen. 
This effect gets  enhanced when the general mass scale
$m_{\tilde{q}}$ in the 
squark mass matrices decreases with the parameter $M_{\rm susy}$.
\begin{figure}[t]
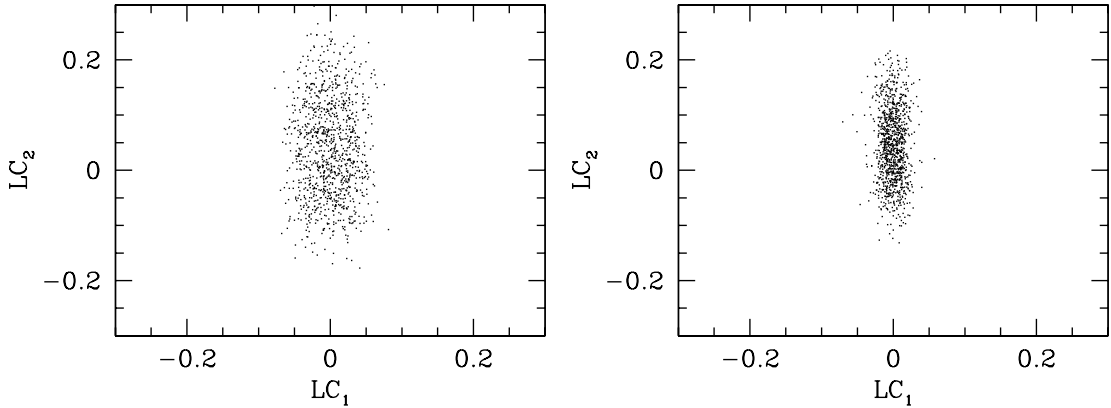

    \begin{center}
    \leavevmode
    \includegraphics[height=6cm,bb= 20 145 560 586]{tanbeta501.epsi}
    \includegraphics[height=6cm,bb= 20 145 560 586]{tanbeta502.epsi}
    \vspace{2ex}
    \caption[f1]{The left frame shows the bounds on $LC_1$ and $LC_2$ 
                 in a $\tan \beta =50$ scenario, which are relatively weak. 
                 In the right frame we put $(F_{d,LR})_{33}=0$. See text.
                 The other parameters are 
                 $\mu=300\,GeV$, $M_{H^-}=300\,GeV$, $M_{\rm{susy}}=500\,GeV$ 
                 and $x=1$. All contributions are switched on.} 
    \label{fig:tanbeta50}
   \end{center} 
\end{figure}
\begin{figure}[t]
    \begin{center}
    \leavevmode
    \includegraphics[height=6cm,bb= 20 145 560 586]{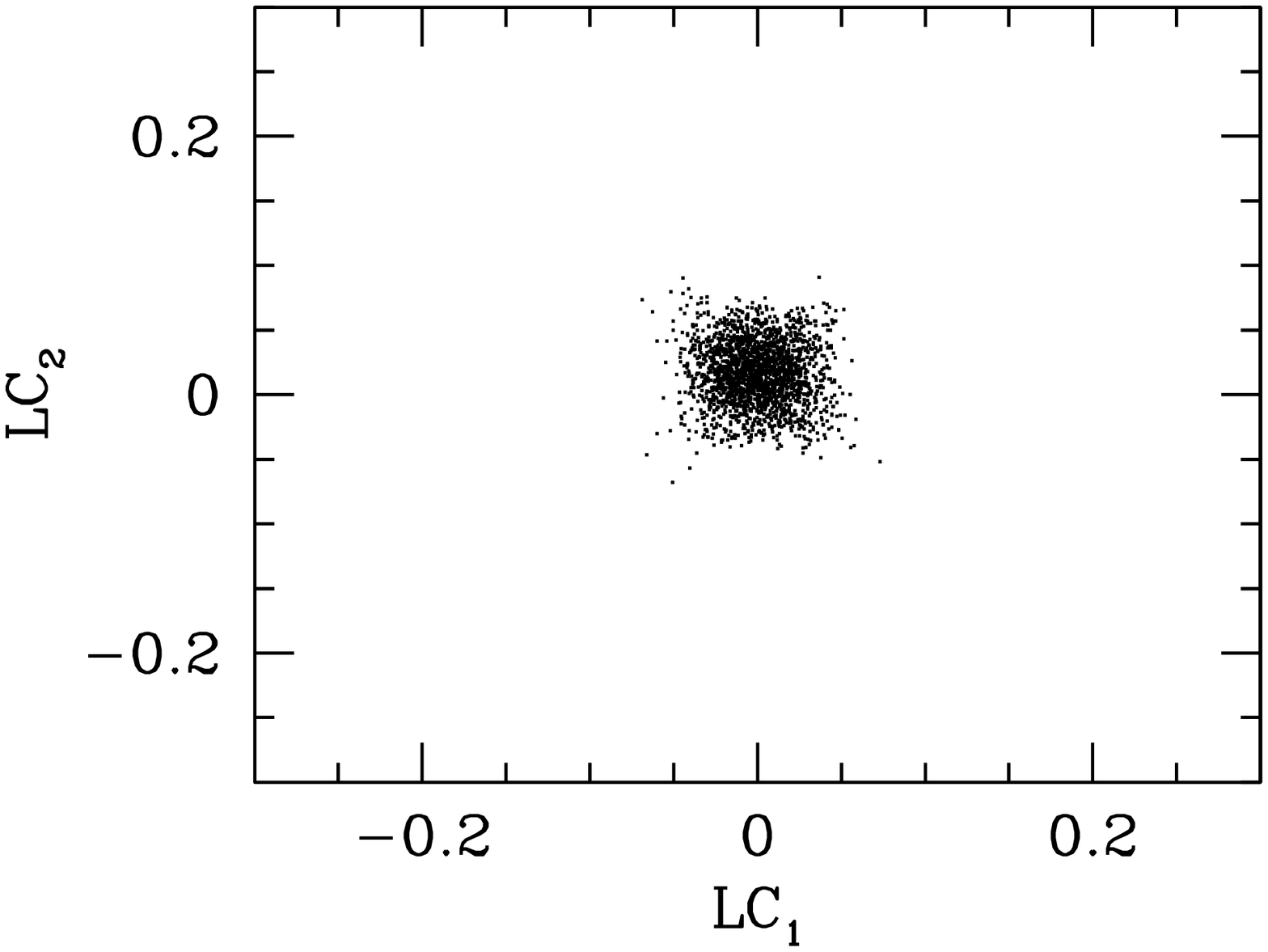}
    \includegraphics[height=6cm,bb= 20 145 560 586]{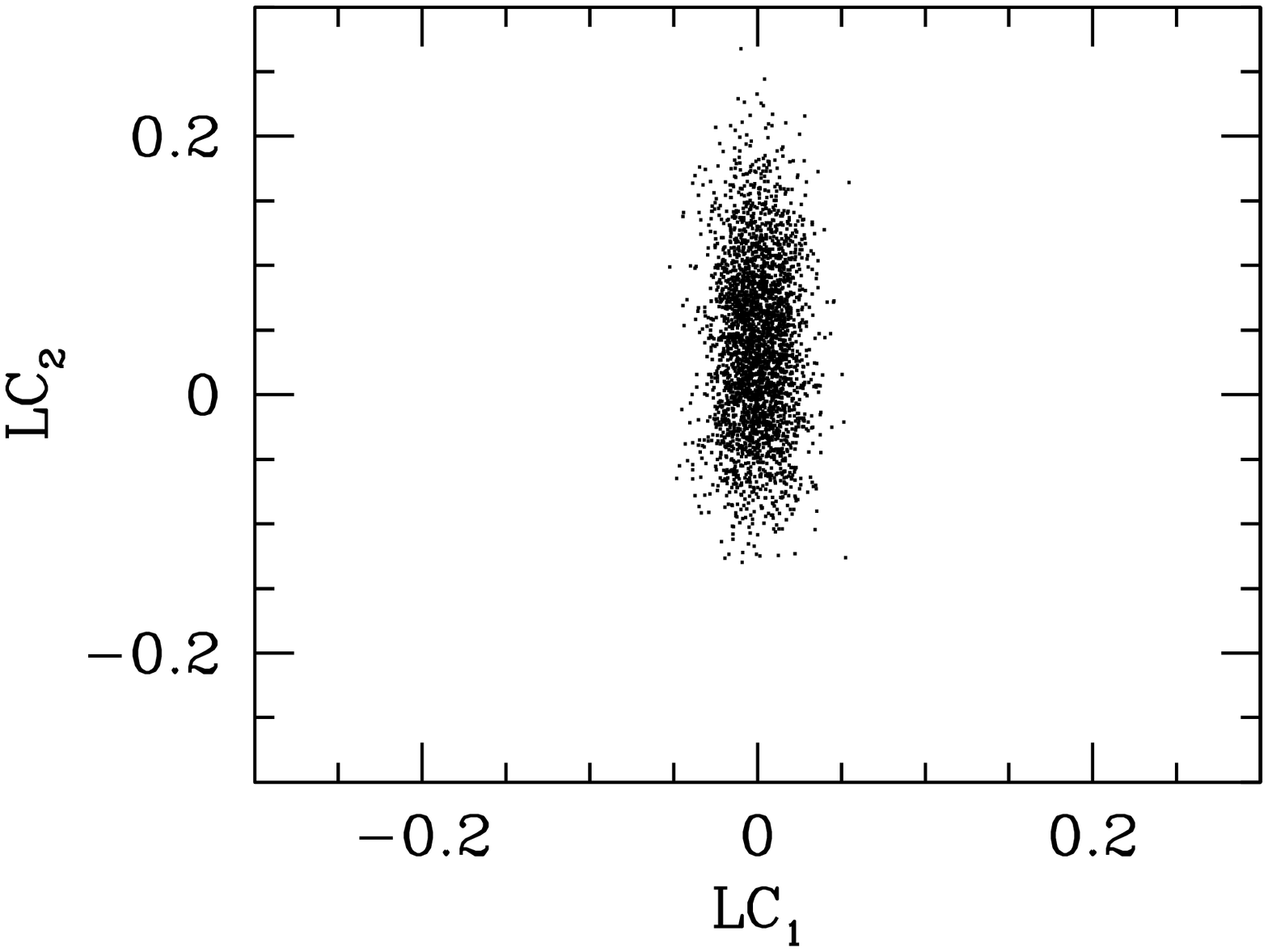}
    \vspace{2ex}
    \caption[f1]{The left frame shows bounds on $LC_1$ and $LC_2$ 
                 for a scenario
                 with  $\mu = 900 \, GeV$ and $\tan \beta = 10$.
                 The right frame shows the corresponding bounds obtained
                 with  $\mu = 150 \, GeV$ and $\tan \beta = 30$. The other 
                 parameters in both frames are chosen to be
                 $M_{H^-}=300 \, GeV$, 
                 $M_{\rm{susy}}=500\, GeV$ 
                 and $x=1$. All contributions
                 are switched on. See text.}
    \label{new150}
   \end{center} 
\end{figure}
There are two main reasons why the bounds get weakened in these scenarios.
First, in the large $\tan \beta$ regime the term 
$(F_{d,LR})_{33}$ gets strongly enhanced because of its proportionality  
to $\tan \beta$ (see (\ref{FFterm})). Particularly, for $\tan \beta = 50$ 
and $M_{{susy}} = 300\, GeV$, the term  $(F_{d,LR})_{33}$ is of the
same magnitude as the diagonal entries of the squark mass matrix. 
Thus,  the contributions to the Wilson coefficients of the five-dimensional
gluino operators (induced by $(F_{d,LR})_{33}$
in combination with
$\delta_{d,LL,23}$ or $\delta_{d,RR,23}$) become important 
enough to weaken the bounds on $LC_1$ and $LC_2$ significantly.
The relative importance of this $F$ term is of course increased  
if the general soft squark mass scale $M_{{\rm susy}}$ is decreased
as can be read off from fig. \ref{allx1}.
Second, within the large $\tan \beta$ regime  
the contributions from charginos get enhanced
and therefore also weaken the bounds on $LC_2$. 

These features are illustrated in more detail in 
fig. \ref{fig:tanbeta50}.
In the first frame we take over the specific scenario 
with $\tan\beta=50$ and $M_{{\rm susy}}=500\, GeV$ from fig. \ref{allx1}.
To show that the term $(F_{d,LR})_{33}$ is indeed 
one of the reasons  for the weakening of the bounds, we present in 
the right frame 
of fig. \ref{fig:tanbeta50} the corresponding scenario 
when $(F_{d,LR})_{33}$ is set to zero. 
We see that we regain better bounds on $LC_1$ and also on $LC_2$.
However, we also  see that the bound on $LC_2$ remains weak. 
This, and the resulting asymmetry, is due to a large 
chargino contribution for $\tan\beta=50$. We recall that there is 
no chargino contribution
to the primed operator which could influence
the bound on $LC_1$. 

We can also explore how the bounds behave if we vary the parameter $\mu$.
Until now we used the value $\mu = 300\, GeV$. 
Because the parameter $(F_{d,LR})_{33}$ is actually proportional to the 
product of $\tan \beta$ and $\mu$ (see eq. (\ref{FFterm})), we conclude 
from the findings above that the bound on $LC_1$ is  unchanged if we 
increase the value of $\mu$ and decrease the value of $\tan \beta$ such 
that the product of both parameters is constant; the bound on
$LC_2$ is then  even stronger because the chargino contribution 
is smaller for increasing $\mu$. Consequently, one finds  a smaller asymmetry 
in the corresponding plots (compare the left frame in fig. \ref{new150}
with the second frame in the second line of fig. \ref{allx1}). 
On the contrary, if one decreases the value of $\mu$ to $\mu = 150\, GeV$,
the bound  on $LC_2$ is weakened and the asymmetry of the plot
is increased as one can read off from the right frame in
fig. \ref{new150}. 

Summing up the second part of our analysis, 
the two simple combinations 
$LC_1$ and $LC_2$ (\ref{deflc1lc2}), consisting  
of elements of the soft parts of the down squark mass matrices,
stay 
stringently bounded over large parts of the 
supersymmetric parameter space, 
excluding the large $\tan \beta$ and the large $\mu$ regime.
We note that these new bounds are in general one order of magnitude weaker 
than the bound on the single off-diagonal element $\delta_{d,LR,23}$, which 
was derived in previous work \cite{GGMS,Masiero2001}
by neglecting any kind of interference effects 
(see e.g. tab. 4 in \cite{Masiero2001} where
the value $1.6 \cdot 10^{-2}$ is given as bound
on  $\delta_{d,LR,23}$ for $x=1$ and $M_{\rm{susy}}=500\, GeV$).

\section{Implications on $\hat{C}_8(\mu_W)$ and $\hat{C}_8'(\mu_W)$}
As mentioned in section \ref{framework}, it is possible to absorb the various
versions of gluonic dipole operators into the SM operator
${\cal O}_8$ and its primed counterpart. The resulting 
effective Wilson coefficients,
denoted by $\hat{C}_8(\mu_W)$ and $\hat{C}_8'(\mu_W)$, read
at the matching scale $\mu_W$:
\begin{eqnarray}
\label{combine}
\hat{C}_8(\mu_W) &=& C_8(\mu_W) - 
\left( C_{8b,\tilde{g}}(\mu_W) +
\frac{1}{m_b(\mu_W)} \, C_{8\tilde{g},\tilde{g}}(\mu_W) \right)
\, \frac{16 \sqrt{2} \, \pi^3 \, 
\alpha_s(\mu_W)}{G_F \, K_{tb} K_{ts}^*} \nonumber \\
\hat{C}_8'(\mu_W) &=& C_8'(\mu_W) - 
\left( C_{8b,\tilde{g}}'(\mu_W) +
\frac{1}{m_b(\mu_W)} \, C_{8\tilde{g},\tilde{g}}'(\mu_W) \right)
\, \frac{16 \sqrt{2} \, \pi^3 \, 
\alpha_s(\mu_W)}{G_F \, K_{tb} K_{ts}^*} \quad .
\end{eqnarray}                   
The coefficients on the r.h.s. of eq. (\ref{combine}) are
given explicitly in section \ref{Wilson} (appendix \ref{ami}).

\begin{figure}[t]
    \begin{center}
    \leavevmode
    \includegraphics[height=6cm,bb= 20 145 560 586]{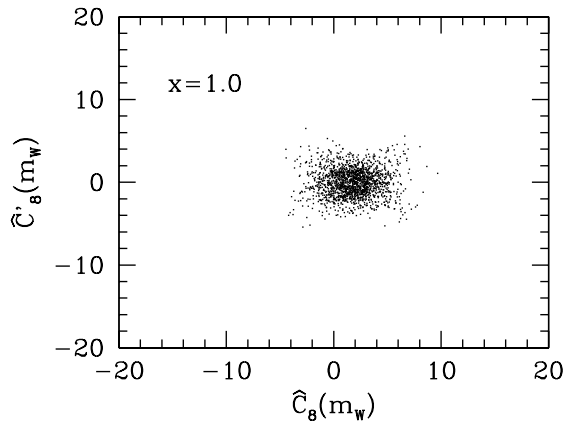}
     \vspace{2ex}
    \caption[f1]{Allowed values for the Wilson coefficients 
             $\hat{C}_8(m_W)$ and
             $\hat{C}_8'(m_W)$ satisfying the experimental constraint on 
             $BR(B \rightarrow X_s\,\gamma)$. 
             The values of the input parameters are
             $\mu=300\,GeV$, $\tan\beta=10$, $M_{H^-}=300\,GeV$, 
             $M_{\rm{susy}}=500\,GeV$ and $x=1$. All contributions are 
             switched on.} 
    \label{fig:c08}
   \end{center} 
\end{figure}

We now investigate the implications on possible
values for the effective Wilson coefficients $\hat{C}_8(\mu_W)$ and 
$\hat{C}_8'(\mu_W)$ 
when taking into account 
the experimental constraints on $B \to X_s \gamma$. The result is shown
in fig. \ref{fig:c08}, for $\mu=300\,GeV$, $M_{H^-} = 300\,GeV$, 
$M_{\rm{susy}} = 500\,GeV$, 
$X_t = 750\,GeV$,
$\tan \beta = 10$ and $x=1$. The soft parameters, encoded in the
$\delta$ quantities, are varied as in fig. \ref{fig:lc1lc2}.
From fig. \ref{fig:c08} we conclude that large deviations from the SM values
for  $\hat{C}_8(\mu_W)$ and $\hat{C}_8'(\mu_W)$ are still possible.
Scenarios in which these Wilson coefficients are enhanced with respect to
the SM gained a lot of attention in the last years.
For a long
time the theoretical predictions for both, the inclusive
semileptonic branching ratio ${\cal B}_{\rm{sl}}$
and the charm multiplicity $n_c$ in $B$-meson
decays were considerably higher than the experimental values
\cite{Bigi_Falk}. 
An attractive
hypothesis, which would move the theoretical predictions
for both observables into the direction favoured by the experiments,
assumed the Wilson coefficients 
$\hat{C}_8(\mu_W)$ and $\hat{C}_8'(\mu_W)$ to be enhanced by new physics
\cite{Kagan}.

After the inclusion of the complete NLL corrections to the decay modes
$b \to c \overline{u} q$ and $b \to c \overline{c} q$ ($q=d,s$) \cite{Bagan},
the theoretical prediction for the central values of the
semileptonic branching ratio
and the charm multiplicity \cite{Sachrajda}
are still somewhat higher than the present measurements
\cite{Japantalk}, but theory and experiment are in agreement
within the errors.
It should be stressed, however, that in the theoretical error estimate the 
renormalization was varied down to 
$m_b/4$. If one only considers the variations down to
$m_b/2$, the theoretical predictions
will have only an marginal overlap with the data. This implies that 
there is still room for enhanced $\hat{C}_8(\mu_W)$ and $\hat{C}_8'(\mu_W)$
\cite{Liniger}.

\section{Summary}
In this paper we have chosen the rare decay 
$B \rightarrow X_s \gamma$ to analyse the importance of 
interference effects for the bounds on the 
parameters in the squark mass matrices within the unconstrained MSSM.
Our analysis, based on a systematic leading
logarithmic (LL) QCD analysis,  mainly explored
the interplay  between the various 
sources of flavour violation and the interference effects of
SM, gluino,
chargino, neutralino, and charged Higgs boson contributions.
Surprisingly, such an analysis did not exist so far.
Unlike previous work, which used the mass insertion approximation, 
we used in our analysis the mass eigenstate formalism, which remains 
valid even when some of the  intergenerational mixing elements are 
large. 

In former analyses no 
correlations between the different sources of flavour violation 
were taken into account.
Following that approach, we found only two $down$-type
squark mass entries to be significantly constrained by the data on 
$B \rightarrow X_s \gamma$: $\delta_{d,LR,23}$ and 
$\delta_{d,RL,23}$. These entries are correlated with the five-dimensional 
dipole operators where the chirality flip is induced by the gluino mass. 
We showed that these  bounds get  destroyed in 
scenarios in which certain off-diagonal elements of the squark mass 
matrices are switched on simultaneously. 

We then systematically explored the interference effects from
all possible contributions and sources of flavour violation
within the unconstrained MSSM.
Accordingly, we switched on all off-diagonal elements $\delta_i$ 
of the squark mass matrices and varied them in the range $\pm 0.5$. 
In addition, we also varied the diagonal elements, but in  smaller 
interval in order to preserve a certain  hierarchy between the off-diagonal
and the diagonal ones. 
In this general scenario we singled out two simple combinations of
elements of the soft part of the down squark mass matrix,  
which stay stringently bounded over large parts of the 
supersymmetric parameter space, excluding the large $\tan \beta$ and the
large $\mu$ regime.
These new bounds are in general one order of magnitude weaker 
than the bound on the single off-diagonal element $\delta_{d,LR,23}$, which 
was derived in previous work \cite{GGMS,Masiero2001} by
neglecting any kind of interference effects.  

Finally, we briefly analysed up to which values SUSY contributions, 
compatible with $B \rightarrow X_s \gamma$, can enhance the 
Wilson 
coefficients $\hat{C}_8 (m_W)$ and $\hat{C}'_8 (m_W)$. 
We found that 
large deviations from the SM values are still possible in our general setting. 
Such scenarios are of particular interest within the phenomenology of
inclusive charmless hadronic $B$ decays.

\acknowledgements
We thank Sven Heinemeyer, Shaaban  Khalil and  Georg Weiglein for discussions.
\appendix
\section{$~$} 
\label{ami}
\subsection{Mixing matrices, interacting Lagrangian}\label{am}

In this  appendix we present our conventions in the mass mixing matrices 
for the relevant particles and in the interacting Lagrangian.  
Besides some specific changes, we follow~\cite{Andreas}: 

\noindent
{\bf Charged Higgs bosons:}\, If we denote the two 
$SU(2)$ Higgs boson doublets appearing in the superpotential by 
\begin{equation}
    H_1=\left(\begin{array}{c}H_1^1\\ 
    H_{1}^{2}\end{array} \right) \, , \hskip 2cm
    H_2=\left(\begin{array}{c}H_2^1\\ 
    H_{2}^{2}\end{array} \right),
\end{equation}
the corresponding mass eigenstates $H_1^{+\,-}$ 
and $H_2^{+\,-}$ of the charged Higgs bosons are given by (see \cite{Rosiek})
\begin{equation}
    \left(\begin{array}{c}H_2^{1\,\ast}\\H_1^2\end{array}\right)
     =\underbrace{\left(\begin{array}{cc}\sin\beta & \cos\beta \\ -\cos\beta & \sin\beta \end{array}\right)}_{Z_E}
       \hskip 0.3cm
      \left(\begin{array}{c}H_2^{-}\\H_1^-\end{array}\right) \, ,
\end{equation}
and similarly for $H_2^+=(H_2^-)^\ast$ and $H_1^+=(H_1^-)^\ast$.\\
In the unitary (physical) gauge, the massless charged fields $H_2^{+\,-}$ 
are absorbed by the $W$ boson. 
One is left with two massive charged Higgs bosons of equal mass.

\noindent 
{\bf Charginos:}\, The charginos $\chi^{ch}_{1/2}$ are a mixture of charged 
    gauginos $\lambda^{\pm}$ and Higgsinos $h_{1}^{-}$ and  $h_{2}^{+}$.
    Defining
    \begin{equation}
    \psi^{+}=\left(\begin{array}{c}-i\lambda^{+}\\ 
    h_{2}^{+}\end{array} \right) \, , \hskip 2cm
    \psi^{-}=\left(\begin{array}{c}-i\lambda^{-}\\ 
    h_{1}^{-}\end{array} \right) \, ,
    \end{equation}
    the mass terms are then
   ${\cal L}_{m}^{ch}=
      -\frac{1}{2}(\psi^{+T}X^{T}\psi^{-}+\psi^{-T}X\psi^{+})+\mbox{h.c.}$, 
    where
    \begin{equation}
    \label{Xmat}
    X=\left(\begin{array}{cc}M_{2}&g_{2}v_{2}\\ g_{2}v_{1}&\mu 
    \end{array}\right).
    \end{equation}
    The two-component charginos $\chi^{\pm}_{i}\;(i=1,2)$ and the 
    four-component charginos $\chi^{ch}_{1/2}$ are then defined as 
    \begin{equation} \renewcommand{\arraystretch}{1.5}
    \begin{array}{llllll}
    \chi_{i}^{+}&=&V_{ij}\psi_{j}^{+},&\chi_{i}^{-}&=&U_{ij}\psi_{j}^{-},\\
    \chi^{ch}_{1}&=&\left(\begin{array}{c} 
    \chi_{1}^{+}\\ 
    \overline{\chi_{1}^{-}}\end{array}\right), &\chi_{2}^{ch}&=& 
    \left(\begin{array}{c}\chi_{2}^{+}\\ 
    \overline{\chi_{2}^{-}}\end{array}\right),
    \end{array}
  \label{M_2}  
  \end{equation}
    where the unitary matrices $U$ and $V$ diagonalize $X$:
    $M_{D}^{ch}=U^{*}XV^{-1}=VX^{\dagger}U^{T}$.
    ${\cal L}^{ch}_{m}$ then becomes
    ${\cal L}^{ch}_m=-M^{ch}_{D\,11}\:\overline{\chi_{1}^{ch}}\chi^{ch}_{1} 
    -M^{ch}_{D\,22}\:\overline{\chi^{ch}_{2}}\chi^{ch}_{2}.$
    $U$ and $V$ can be found by observing that\\
    $M^{ch\,2}_{D}=VX^{T}XV^{-1}=U^{*}XX^{T}U^{*-1}.$
    They are not fixed completely by these conditions. The freedom 
    can 
    be used to arrange the elements of $M^{ch}_{D}$ to be positive: 
    If 
    the $i^{th}$ eigenvalue of $M^{ch}_{D}$ is negative, simply 
multiply 
    the $i^{\rm th}$ row of $V$ with $-1$.

\noindent
{\bf  Neutralinos:}\,  The neutralinos are 
linear combinations of the gauginos $\lambda'$ 
and  $\lambda_{3}$ and the neutral Higgsinos $h_{1}^{0}$ and $h_{2}^{0}$. 
 If we define
 \begin{equation}
 \psi^{0}=\left(\begin{array}{c} -i\lambda'\\-i\lambda_{3}\\h_{1}^{0} 
 \\h_{2}^{0}\end{array}\right) \, ,
 \end{equation}
 the neutralino mass term reads
 ${\cal L}^{0}_{m}=-\frac{1}{2}\psi^{0T}Y\psi^{0}+\mbox{h.c.},$
 where
 \begin{equation}
 Y=\left(\begin{array}{cccc}
 M_{1}&0&-\frac{g_{1}v_{1}}{\sqrt{2}}&\frac{g_{1}v_{2}}{\sqrt{2}}\\
 0&M_{2}&\frac{g_{2}v_{1}}{\sqrt{2}}&-\frac{g_{2}v_{2}}{\sqrt{2}}\\
 -\frac{g_{1}v_{1}}{\sqrt{2}}&\frac{g_{2}v_{1}}{\sqrt{2}}&0&-\mu\\
 \frac{g_{1}v_{2}}{\sqrt{2}}&-\frac{g_{2}v_{2}}{\sqrt{2}}&-\mu&0
 \end{array} \right).
 \end{equation}
 Two- and four-component neutralinos must be defined as
 \begin{equation} \renewcommand{\arraystretch}{1.5}
 \begin{array}{l}
     \tilde{\chi}^{0}_{i}=N_{ij}\psi^{0}_{j} \, , \hskip 1.5cm (i=1,\ldots 
,4)\\
     \chi^{0}_{i}=\left(\begin{array}{c}\tilde{\chi}^{0}_{i} \\ 
     \overline{\tilde \chi^{0}_{i}} \end{array}\right) \, .
 \end{array}
\end{equation} 
To diagonalize the mass matrix, $N$ must obey
$N_{D}=N^{*}YN^{-1}$,
where $N_{D}$ is a diagonal matrix.
$N$ can be found, using the property
$N_{D}^{2}=NY^{\dagger}YN^{-1}.$
The eigenvalues and eigenvectors are found numerically. Possible 
negative entries in $N_{D}$ are turned positive by multiplying the 
corresponding row of $N$ by a factor of $i$.

\noindent
{\bf Quarks:} \, 
The situation in the quark sector is in almost complete 
analogy to that of the SM. The quarks get their masses from the 
Yukawa 
potential when the Higgs bosons acquire a vacuum expectation value. We 
define the mass eigenstates by
\begin{equation}
\begin{array}{ll}
    u_{Li}^{(m)}=U^{L}_{ij}u_{Lj},&u_{Ri}^{(m)}=U^{R}_{ij}u_{Rj},\\
    d_{Li}^{(m)}=D^{L}_{ij}d_{Lj},&d_{Ri}^{(m)}=D^{R}_{ij}d_{Rj}.
\end{array}
\end{equation} 
The mixing matrices must satisfy ($i=1,2,3$)
\begin{equation}
\begin{array}{rlllll}
    
D^{R}\lambda^{dT}D^{L\dagger}&=&\lambda^{d}_{D}&=&\mbox{diag}\left( 
    \frac{m_{di}}{v_{1}}\right),&\\
    U^{R}\lambda^{uT}U^{L\dagger}&=&\lambda^{u}_{D}&=&\mbox{diag}\left( 
    \frac{m_{ui}}{v_{2}}\right),&
\end{array}
\end{equation}
where
\begin{equation}
\begin{array}{lll}
v_1&=&\sqrt{2}\frac{m_W}{g_2}\cos \beta,\\
v_2&=&\sqrt{2}\frac{m_W}{g_2}\sin \beta.
\end{array}
\end{equation}
As can be seen, the eigenvalues of $\lambda^{u}$ and $\lambda^{d}$ 
are 
fixed by the quark masses and the minimum of the Higgs potential.
In the SM, the only observable effect of the mixing is encoded in the 
CKM matrix $K=U^{L}D^{L\dagger}$, appearing in the  
charged current. Therefore it is possible and convenient to set 
$D^{L}=D^{R}=U^{R}=\openone\,\, (\Rightarrow U^{L}=K)$.
To be more precise, $\lambda^{d}$ and $\lambda^e$ are chosen 
to be diagonal and 
$\lambda^{u}=\mbox{diag}\left(\frac{m_{ui}}{v_{2}}\right) K^{T}$.
Although in our theory the mixing matrices appear in all kinds of 
combinations, we adopt this convention here, emphasizing that it is a 
\emph{choice} made just for convenience. 
An  underlying theory should fix the values of $\lambda^{u}$ and $\lambda^{d}$ 
at some (high) scale. Note that in the main text we 
neglect the superscript $m$ for the 
mass eigenstates.

\noindent
{\bf  Squarks:}\, 
If supersymmetry were not broken, squarks would be 
rotated to 
their mass basis with the help of the same matrices as their 
fermionic partners. In a more realistic setting  we need 
to introduce a further set of unitary rotation matrices. The notation 
must be set up carefully because the mass eigenstates of squarks and 
sleptons are linear combinations of the partners of \emph{left-} and 
\emph{right-}handed partners of the corresponding fermions. The exact
form of the mass matrices and the notation for the corresponding
diagonalization matrices can be found in section \ref{SMM}.\\

\noindent
{\bf Interaction Lagrangian:}\, In order to fix further conventions 
we quote the relevant parts of the interaction Lagrangian: 
\begin{itemize}
\item{Charged Higgs boson-quark-quark}
\begin{eqnarray}\label{Higgs}
  {\cal L}_{qqH}&=&(\lambda_D^d)_{m\ell}K_{mk}\overline{d_\ell}P_L u_{k}(Z_E)_{2\,n}H^-_i\,
                  +\,(\lambda_D^d)_{m\ell}K_{mk}^\ast\overline{u_k}P_R d_{\ell}(Z_E)_{2\,n}^\ast(H^-_i)^\ast 
                  \nonumber \\
         & + & (\lambda_D^u)_{m\ell}\overline{u_\ell}P_L d_{m}(Z_E)_{1\,n}^\ast(H^-_i)^\ast\,
                +\,\lambda_D^u)_{m\ell}\overline{d_m}P_R u_{\ell}(Z_E)_{1\,n}H^-_i.
\end{eqnarray}
Note that in our basis, the terms proportional to the 
$\lambda_D^d$ always come together with the CKM matrix $K$, 
while the $\lambda_D^u$ terms do not.
\item{Squark-quark-chargino}
\begin{eqnarray}\label{QSQCH}
   {\cal L}_{\tilde qq\chi^{ch}}&=& \tilde u_{j}\overline{d_{i}}
\left[A^d_{ij\ell}P_L +B^d_{ij\ell}P_R \right]\chi^{ch\;c}_\ell 
+\tilde
u_i^\dagger
\overline{\chi^{ch\;c}_\ell} \left[A^{d\dagger}_{ij\ell}P_R
+B^{d\dagger}_{ij\ell}P_L\right]d_j \, , 
\end{eqnarray}
where 
$A^d_{ij\ell}=(\lambda^d_D\Gamma^{\dagger}_{UL})_{ij}U^*_{\ell 2}$,
$B^d_{ij\ell}=(K^\dagger\lambda^u_D\Gamma^{\dagger}_{UR})_{ij}V_{\ell 
2}
-g_2\Gamma^{\dagger}_{ULij} V_{\ell 1}$,
$P_{L/R}=\frac{1}{2}(1\mp\gamma^5)$
and $\chi^{ch\;c}_\ell$ denotes the charge-conjugated field.
\item{Squark-quark-neutralino}
\begin{eqnarray}\label{QSQNE}
   {\cal L}_{\tilde qq\chi^0}&=&-\tilde 
d_j\overline{d_i}\left[C^d_{ij\ell}P_L
+D^d_{ij\ell}P_R\right]\chi^0_\ell -\tilde
d^\dagger_i\overline{\chi^0_\ell}\left[C^{d\dagger}_{ij\ell}P_R
+D^{d\dagger}_{ij\ell}P_L\right]d_j \, ,
\end{eqnarray}
where
$C^d_{ij\ell}=(\lambda^d_D\Gamma^{\dagger}_{DL})_{ij}N^*_{\ell 3}
-\sqrt{2}g_1Q_d\Gamma^{\dagger}_{DRij}N^*_{\ell 1}$,\\
$D^d_{ij\ell}=(\lambda^d_D\Gamma^{\dagger}_{DR})_{ij}N_{\ell 3}$
$+\frac{1}{\sqrt{2}}\Gamma^{\dagger}_{DLij} ((2Q_d+1)g_1N_{\ell 1}-g_2 
N_{\ell 2})$.
\end{itemize}
\subsection{Wilson coefficients}\label{Wilson}
We recall the Wilson coefficients 
at the matching scale $\mu_W$. The non-vanishing Wilson coefficients
for the SM  are, at leading order in $\alpha_s$ \, 
($x_{tw} \equiv m_t^2/m_W^2$) :
\begin{eqnarray}
 C_{2SM}(\mu_W)  & = & 1
\nonumber \\[1.5ex]
 C_{7SM}(\mu_W)  & = & \frac{x_{tw}}{24\,(x_{tw}-1)^4} \, \left(
 {-8x_{tw}^3+3x_{tw}^2+12x_{tw}-7+(18x_{tw}^2-12x_{tw}) \ln x_{tw}} \right) 
\nonumber \\[1.5ex]    
 C_{8SM}(\mu_W)  & = & \frac{x_{tw}}{\ 8\,(x_{tw}-1)^4} \, \left(
 {-x_{tw}^3+6x_{tw}^2-3x_{tw}-2-6x_{tw} \ln x_{tw}}      \right) 
\,.
\label{wclosm}                                          
\end{eqnarray}
The contributions from charginos, neutralinos and charged Higgs bosons 
match onto the (chromo)magnetic operators of the SM and the corresponding 
primed operators, which differ from the SM ones 
only by their chirality structure. 
The corresponding Wilson coefficients become somewhat involved
\cite{Andreas} as they include
many mixing matrices, whose definitions were  given in appendix A1.
One gets (using the abbreviation 
$V \doteq (4 G_F \, K_{tb} K_{ts}^*)/\sqrt{2}$) 
\begin{eqnarray}
C_{7}(\mu_W) & = & C_{7\mathit{SM}}(\mu_W) \nonumber \\
 & - & \frac{1}{2} \, [\cot^2\beta\,
 x_{tH}(Q_{u}F_1(x_{tH})+F_2(x_{tH}))+ \nonumber \\
 &&x_{tH}(Q_uF_3(x_{tH})+F_4(x_{tH}))] +
 \nonumber \\
 & + &\frac{1}{2V}\frac{1}{m^2_{\tilde
 u_j}}B^d_{2j\ell}B^{d*}_{3j\ell}\left[F_1(x_{\chi^{ch}_\ell u_j})+
Q_{u}F_2(x_{\chi^{ch}_{\ell u_j}})\right] \nonumber\\
&+&\frac{1}{2V}\frac{1}{m^2_{\tilde
 u_j}}\frac{m_{\chi^{ch}_\ell}}{m_b}B^d_{2j\ell}A^{d*}_{3j\ell}
\left[F_3(x_{\chi^{ch}_\ell
  u_j})+Q_u F_4(x_{\chi^{ch}_\ell
  u_j})\right] \nonumber\\
&+& \frac{Q_d}{2V}\frac{1}{m^2_{\tilde
 d_j}}\left[D^d_{2j\ell}D^{d*}_{3j\ell}F_2(x_{\chi^0_\ell \tilde
 d_j})+\frac{m_{\chi^0_\ell}}{m_b}D^d_{2j\ell} 
C^{d*}_{3j\ell}F_4(x_{\chi^0_\ell
 \tilde d_j})\right]\label{C7}\\
C_{8}(\mu_W) & = & C_{8\mathit{SM}}(\mu_W) \nonumber \\
 & - & \frac{1}{2} \, [\cot^2\beta\,
x_{tH}F_1(x_{tH})+x_{tH}F_3(x_{tH})] \nonumber \\
& + & \frac{1}{2V}\frac{1}{m^2_{\tilde
u_j}}\left[B^d_{2j\ell}B^{d*}_{3j\ell}F_2(x_{\chi^{ch}_\ell 
u_j})
+\frac{m_{\chi^{ch}_\ell}}{m_b}B^d_{2j\ell}A^{d*}_{3j\ell}
F_4(x_{\chi^{ch}_\ell
u_j})
\right] \nonumber \\
& + & \frac{1}{2V}\frac{1}{m^2_{\tilde
d_j}}\left[D^d_{2j\ell}D^{d*}_{3j\ell}F_2(x_{\chi^0_\ell 
d_j})
+\frac{m_{\chi^0}}{m_b}D^d_{2j\ell}C^{d*}_{3j\ell}F_4(x_{\chi^0_\ell 
d_j})
\right]\label{C8} \\
C_{7}^{\prime}(\mu_W) & = &
-\frac{1}{2} \, \frac{m_sm_b}{m_t^2}\tan^2\beta\,
 x_{tH}(Q_{u}F_1(x_{tH})+F_2(x_{tH})) \nonumber \\
 & + &\frac{1}{2V}\frac{1}{m^2_{\tilde
 u_j}}A^d_{2j\ell}A^{d*}_{3j\ell}\left[F_1(x_{\chi^{ch}_\ell 
u_j})+Q_{u}F_2(x_{\chi^{ch}_\ell 
 u_j})\right] \nonumber\\
&+&\frac{1}{2V}\frac{1}{m^2_{\tilde
u_j}}
\frac{m_{\chi^{ch}_\ell}}{m_b}A^d_{2j\ell}B^{d*}_{3j\ell}
\left[F_3(x_{\chi^{ch}_\ell
 u_j})+Q_u F_4(x_{\chi^{ch}_\ell
  u_j})\right] \nonumber \\
& + & \frac{Q_d}{2V}\frac{1}{m^2_{\tilde
 d_j}}\left[C^d_{2j\ell}C^{d*}_{3j\ell}F_2(x_{\chi^0_\ell 
 d_j})
+\frac{m_{\chi^0_\ell}}{m_b}C^d_{2j\ell} 
D^{d*}_{3j\ell}F_4(x_{\chi^0_\ell
 d_j})
\right]\label{C7'} \\
C_{8}^{\prime}(\mu_W) & = &
 -\frac{1}{2} \, \frac{m_sm_b}{m_t^2}\tan^2\beta\,
x_{tH}F_1(x_{tH}) \nonumber \\
& + & \frac{1}{2V}\frac{1}{m^2_{\tilde
u_j}}\left[A^d_{2j\ell}A^{d*}_{3j\ell}F_2(x_{\chi^{ch}_\ell 
u_j})
+\frac{m_{\chi^{ch}_\ell}}{m_b}A^d_{2j\ell}B^{d*}_{3j\ell}
F_4(x_{\chi^{ch}_\ell
u_j})
\right] \nonumber \\
& + & \frac{1}{2V}\frac{1}{m^2_{\tilde
d_j}}\left[C^d_{2j\ell}C^{d*}_{3j\ell}F_2(x_{\chi^0_\ell 
d_j})
+\frac{m_{\chi^0_\ell}}{m_b}C^d_{2j\ell}D^{d*}_{3j\ell}
F_4(x_{\chi^0_\ell d_j})
\right] \, ,
\label{C8'}
\end{eqnarray}
where $Q_u=2/3$ and $Q_d=-1/3$. 
We kept the charged Higgs boson
contribution to the primed operators since they are proportional
to $\tan^2 \beta$ which could compensate the $m_s/m_t$ suppression.
The functions $F_i(x)$ are defined at the end of this section.
Although the Wilson coefficients $C_7'(\mu_b)$
and  $C_8'(\mu_b)$ of the primed 
operators are usually small, we retain them in our analysis.

Among the coefficients arising from the virtual exchange of a
gluino, the most important ones are those associated with the
(chromo)magnetic operators:
 \begin{eqnarray}
C_{7b,\tilde{g}}(\mu_W)              & = &
\ \ 
-\frac{Q_d}{16 \pi^2} \, \frac{4}{3} 
 \sum_{k=1} ^6 \frac{1}{m_{\tilde{d}_k}^2} 
\left( \Gamma_{DL}^{kb} \, \Gamma_{DL}^{\ast\,ks} \right)
 F_2(x_{gd_k}) \, ,
                                     \nonumber \\
C_{7\tilde{g},\tilde{g}}(\mu_W)      & = & 
 m_{\tilde g}\,
 \frac{Q_d}{16 \pi^2} \, \frac{4}{3}
 \sum_{k=1} ^6 \frac{1}{m_{\tilde{d}_k}^2} 
\left( \Gamma_{DR}^{kb} \, \Gamma_{DL}^{\ast\,ks} \right)
 F_4(x_{gd_k})\, ,            
\nonumber \\
C_{8b,\tilde{g}}(\mu_W)              & = &
\ \ 
-\frac{1}{16 \pi^2} 
 \sum_{k=1} ^6 \frac{1}{m_{\tilde{d}_k}^2} 
\left( \Gamma_{DL}^{kb} \, \Gamma_{DL}^{\ast\,ks} \right) \,  
\left[ - \frac{1}{6} F_2(x_{gd_k})
       - \frac{3}{2} F_1(x_{gd_k}) 
\right] \, ,
                                     \nonumber \\
C_{8\tilde{g},\tilde{g}}(\mu_W)      & = & 
 m_{\tilde g}\, 
 \frac{1}{16 \pi^2} 
 \sum_{k=1} ^6 \frac{1}{m_{\tilde{d}_k}^2} 
\left( \Gamma_{DR}^{kb} \, \Gamma_{DL}^{\ast\,ks} \right) \,            
\left[- \frac{1}{6} F_4(x_{gd_k})
      - \frac{3}{2} F_3(x_{gd_k}) 
\right] \,.
\label{glgl}                
\end{eqnarray}
Note that the coefficients
$C_{7\tilde{g},\tilde{g}}(\mu_W)$ and
$C_{8\tilde{g},\tilde{g}}(\mu_W)$ are of higher dimensionality to
compensate the lower dimensionality of the corresponding operators.
The ratios $x_{gd_k}$ are defined as 
$x_{gd_k} \equiv m_{\tilde g}^2/m_{\tilde{d}_k}^2$.
The Wilson coefficients of the
corresponding primed operators (which are not small
numerically) are obtained through the interchange
$\Gamma_{DR}^{ij} \leftrightarrow \Gamma_{DL}^{ij} $ in
eqs.~(\ref{glgl}).  
For the Wilson coefficients of the scalar/tensorial four-quark operators
we refer  to \cite{BGHW}.
Finally, we define the functions $F_i$ appearing in the Wilson coefficients
listed above:
\label{functions}
\begin{eqnarray}
 F_1(x) &\quad = \quad &  \frac{1}{ 12\, (\!x-1)^4}
  \left( x^3 -6x^2 +3x +2 +6x\log x\right)  \, ,
  \nonumber \\
                                                 &  &  \nonumber \\
 F_2(x) & \quad = \quad & \frac{1}{ 12\, (\!x-1)^4} 
  \left(2x^3 +3x^2 -6x +1 -6x^2\log x\right) \, , 
  \nonumber \\
                                                 &  &  \nonumber \\
 F_3(x) & \quad = \quad & \frac{1}{\phantom{1} 2\, (\!x-1)^3} 
  \left( x^2 -4x +3 +2\log x\right) \, ,
  \nonumber \\
                                                 &  & \nonumber \\
 F_4(x) & \quad = \quad & \frac{1}{ \phantom{1} 2\, (\!x-1)^3}
  \left( x^2 -1 -2x\log x\right) \, .  
\label{loopfunc}
\end{eqnarray}

\end{document}